\documentclass[mathleft
]{an}
\usepackage{graphicx}
\usepackage{times}
\usepackage{natbib}
\bibpunct{(}{)}{;}{a}{}{}
\def\elt{E-ELT}
\def\mic{$\mu\mathrm{m}$}

\def\lesssim{\ifmmode\stackrel{<}{_{\sim}}\else$\stackrel{<}{_{\sim}}$\fi}
\def\gtrsim{\ifmmode\stackrel{>}{_{\sim}}\else$\stackrel{>}{_{\sim}}$\fi}

\begin{document}

\Pagespan{1}{}
\Yearpublication{2006}%
\Yearsubmission{2005}%
\Month{11}%
\Volume{999}%
\Issue{88}%

\title{Prospects of Stellar Abundance Studies from Near-IR
Spectra Observed with the E-ELT}

\author{N. Ryde\inst{1}\fnmsep\thanks{
  \email{ryde@astro.lu.se}\newline}
}
\titlerunning{Near-IR abundance studies with the E-ELT}
\authorrunning{N. Ryde}
\institute{Lund Observatory, Box 43, SE-22100 Lund, Sweden}

\received{30 May 2005}
\accepted{11 Nov 2005}
\publonline{later}


\abstract{%
In 2006 ESO Council authorized a Phase B study of  a European 
AO-telescope with a 42~m segmented primary  with a 5-mirror design, the E-ELT.
Several reports and working groups 
have already presented science cases for an E-ELT,
specifically exploiting the new capabilities of such a large
telescope. One of the aims of the design has been to find a balance in the
performances between an E-ELT and the James Webb Space Telescope,
JWST. Apart from the larger photon-collecting area, the strengths of
the former is the higher attainable spatial and spectral
resolutions. The \elt\ AO system will have an optimal performance in the near-IR, which makes it
specially advantageous. High-resolution spectroscopy in the near-infrared has,
however, not been discussed much. This paper aims at filling that
gap, by specifically
discussing spectroscopy of stellar  (mainly red giant), photospheric abundances. Based on studies in the literature of stellar abundances, at the needed medium
to high spectral resolutions in the near-infrared  ($0.8-2.4\,\mu$m), 
I will try to extrapolate published results to the performance of the \elt\ and
explore what could be done at the E-ELT in this field. A discussion
on what instrument characteristics that would be needed for stellar
abundance analyses in the
near-IR will be given.}

\maketitle

\section{Introduction} \label{start}

The objective for P. Connes when he in 1967 set up his Fourier Transform Spectrometer (FTS) for 
astronomical,\linebreak near-IR spectroscopy, was the investigation of planets\linebreak \citep{connes}. However, a few stars were observed and among the first stellar results was a measurement of the $^{12}$C/$^{13}$C ratio for Betelgeuse \citep{spinrad}.  In the seventies and eighties, most high-resolution IR studies of stars were devoted to studying bright IR stars, such as the dynamics in mira stars and isotopic ratios in M giants, carbon stars, and mira stars, see e.g. \citet{maillard}, all with FTSs.  An early investigation of abundances of elements in stars (C, N, and O) was done by \citet{lambert:84}  for Betelguese. Later, the CNO elements and the isotopic ratio of carbon were determined for 30 Galacitc carbon stars by \citet{lambert:86}. A large step forward in order to study abundances in fainter stars too, was taken with the development of cryogenic echelle spectrometers.

Spectroscopic studies of abundances in red stars
benefit from being performed in the near-infrared at medium or
preferably at high spectral resolution. As we will see, a large range of questions
can be investigated, from the formation and evolution of the Milky
Way  and the Magellanic Clouds, to stellar structure and the evolution
of stars of very different metallicites.  But also the difficult
task of observing and modelling the coolest dwarf stars, the M, L,
and T dwarfs and brown dwarfs can now be done with high-resolution,
near-infrared spectrometers. The detailed study of AGB
stars\footnote{AGB stars - the Asymptotic Giant Branch Stars - are
bright red giants on their second ascent on the giant branch} and
their winds can be best analysed in the near-IR where most of their
flux is emitted. High-resolution is needed to disentangle the
molecular spectra, which contain much information about the star. Cool M dwarfs have a large amount of lines in their spectra and the continuum is difficult to define. For such a case, a very high resolution ($R>60,000$) is really needed. The only way to analyses these type of spectra is by calculating detailed synthetic spectra.

With a medium-to-high spectral resolution spectrometer for the
near-IR on \elt, we will be able to expand our views and study other
galaxies in detail, their chemical evolution and different
structures. We will be able to start addressing detailed questions
on the formation and evolution of other galaxies and start comparing
different galaxies of various kinds. The investigation will provide
unique empirical pieces of evidence for the fundamental, and not
understood question of how galaxies form and evolve. Furthermore,
fundamental questions concerning nucleosynthesis and yields from
stars can be investigated by spectroscopically exploring stars in
different environments. The stellar structures and the evolution of
red giants of different metallicities can be studied. Fainter
objects, such as cool dwarf stars and brown dwarfs, will be surveyed
opening up this field. Metal-poor stars that need  spectra with a high signal-to-noise (SNR)
will be studied in detail and in larger numbers, investigating the
large C- and N-bearing molecular contents in these stars. Hence, the
\elt\ era will provide many exiting scientific questions to address.

Thus, studying stellar abundances is of interest for many reasons. For
example, assembling abundances of a variety of elements from an
ensemble of stars ranging over different metallicities in galaxies or a component of a 
galaxy, will determine
abundance trends and spatial composition gradients. This will
eventually provide us with information on metallicity
distributions, the star-formation rates,\linebreak initial-mass functions, and
the distribution of populations within the systems. Furthermore, the
ages of the populations, time-scales of the enrichment of the
systems and the merger history of the galaxy structures can be
investigated. All this will place constraints on how the systems
formed and evolved, whether merger events dominated
or whether the systems evolved slowly without interactions.

Traditionally, abundance analyses of stellar populations have been
done in the UV/optical wavelength region,\footnote{The CODEX
spectrograph \citep{codex} studied for the \elt\ will address this
need. It is planned for the $\lambda=0.4-0.8\,\mu$m range
\citep{stc:66}}\linebreak preferably using high-resolution, cross-dispersed,
echelle\linebreak spectrometers observing main-sequence stars and giants, especially F, G, and K stars. Red giants are very luminous stars\footnote{A red
supergiant of an effective temperature of $3600$~K like Betelgeuse
has a maximum flux, $F_\lambda$, around $700-1000$~nm and the red
giant R Dor ($T_{\mathrm{eff}}=3000$ K) shines strongest in the
range of $1-1.5\,\mu$m.}  and are
therefore useful as probes for studies in regions such as the
Galactic bulge or external galaxies, which may not be readily
accessed by observing
dwarfs or sub-giants. It should also be noted that red giants are responsible for most of the light
in other galaxies. 
For instance, with the \elt\ stellar abundance investigations of
giants as far away as in the most distant galaxies in the Local
Group will be possible. Late-type giants are brightest in the
near-IR with a sharp decline in flux towards the blue, implying that
near-infrared\footnote{The J, H, and K bands are the most effective
regions, whereas the L and M bands in the thermal infrared have a
large telluric background brightness to overcome. However, the
fundamental band of CO at 4.6 \mic\ might still be of interest.} spectroscopy would be more suitable for a detailed
determination of abundances of these stars. A further advantage when
it comes to the \elt, is the performance of its adaptive optics (AO)
system which will be optimal in the near-IR. This means that higher
spatial resolutions will be achievable than in the optical and
shorter observing times are required compared to not using AO. Thus,
an abundance analysis of red-giant stars in the near-infrared would
be rewarding. Fortunately, sensitive medium and high-resolution
spectroscopy has become possible in the near-IR due to the
development of IR detector technology. The art of determining
chemical abundances in cool stars in the near-IR has benefited
strongly from the realization of sensitive, cryogenic echelle spectrographs capable of
providing high-resolution, near-infrared spectra, such as the
Phoenix \citep[ started operation in the late 90:s at KPNO and
later at Gemini South Observatory]{phoenix} and the CRIRES (at ESO) spectrographs \citep{crires1,crires2}. 

Late-type dwarfs are useful probes of the chemo-\linebreak dynamic evolution of
stellar populations, since they span a range in age, and their
surface abundances reflect the composition of the gas clouds from
which they were once\linebreak formed and as such trace the star-formation
history and enrichment of the interstellar medium. This is encoded
in the stellar elemental abundances.
Late-type giants have evolved from the low-mass stars, the most common 
stars in galaxies. They too span a
range of ages and the warmer ones should have mostly unaffected
surface compositions, with the exception of C and N, which can not
be assumed to be pre-stellar. C+N is, however, assumed to be
conserved for stars on their first ascent up the giant branch.  AGB stars are more complicated and their surface abundances are affected by internal processes. If chemical abundances in giants could be measured as
accurately as for solar-type stars, and be understood theoretically,
e.g. in terms of the modifications of the initial stellar abundances
by the individual stars, their value as probes would be even
greater. To a considerable degree, such further understanding relies
on more systematic abundance measurements. 
Age determinations are, however,  difficult and a problem. This review will emphasize on  the use of 
stellar abundances studies of red giants, due to their brightness.

\section{Studies of element abundances based on
near-IR spectroscopy of late-type stars}

Near-IR spectroscopy fits well into the emerging realization
that the near-IR is a preferred wavelength region for many
 science cases for the \elt. There is an emphasis on the
near-IR not only due to other science drivers, but also since the
high spatial resolution of the \elt\ will first be realized in the
IR, with the Adaptive Optics (AO) system having an optimal
performance there. For example, \cite{olsen:2006} show the promise of ground-based AO systems
by using the NIRI camera on the Gemini North telescope to retrieve colour-magnitude diagrams of resolved stars in the crowded regions in the 
bulge and disk of M31, the nearest massive spiral galaxy in the Local Group of galaxies. Based on H and K band observations they could retrieve star-formation histories for these populations and could  show that there does not seem to be an age difference between them.
Most stellar population work is confusion-limited rather than\linebreak
sky-background or photon-limited, which means that 
we need the high spatial resolution achievable in the near-IR.
For example, in the K band the diffraction limit will be less than
$0.015"$ for a 42-meter telescope, which means that it will be
possible to resolve stars lying 1/20 pc apart in projected distance in the Andromeda galaxy, cf. Table \ref{mag}.





\subsection{The virtues of exploring stellar abundances  by near-infrared spectroscopy}

There is a number of advantages of observing stellar abundances of
late-type stars in the near-infrared\linebreak wavelength region. Below a few aspects are mentioned and
highlighted. For instance, the near-IR is preferred for studies of dust-obscured
regions since IR radiation penetrates gas and dust much better than
optical radiation. The galactic  bulge is an example of a region
which is hidden behind massive amounts of gas and dust. The bulge is
the last major component of the Milky Way which is still essentially
unexplored, mostly due to this fact. Furthermore, cool stars can have a fair amount of molecular lines and might have
messy optical spectra. Near-IR absorption spectra are less
crowded with lines and fewer lines are blended, compared to
wavelength regions in the optical spectral window. It is thus easier
to find portions of the spectrum which can be used to define
a continuum, which is so very important in an abundance analysis. This
reduces the uncertainties significantly and makes the exploration
easier and more accurate \citep{ryde_munchen_review}. Especially, the study of mira stars and carbon stars
benefits largely when going to the near-IR.
In fact, also
the number of atomic and ionic lines is much smaller in the near-IR
than in the ultraviolet.
This is, however,
also a general drawback of the infrared spectral region - the number
of useful atomic and ionic lines is relatively limited in practice.
However, it should be realized that there are enough of suitable metallic
lines in the infrared for many purposes, especially from the diagnostically important
$\alpha$ elements (such as Mg, Si, Ca, S, and Ti).

Molecular lines are ubiquitous in the IR which
implies the possibility to more reliable abundances. Note, however, that the abundance of molecules are often quite temperature sensitive.
Obviously, it is an advantage to have many lines of different 
strengths from the same molecule to minimize measuring uncertainties
and systematic effects. Furthermore, several diagnostics for the same element might 
exist. For instance, in order to
determine the carbon abundance it is possible to measure lines from
CH, CO, C$_2$, and CI. The near-IR also has the virtue that only
that region offers all indicators necessary for an accurate
determination of the important C-N-O molecular and atomic equilibria
in the atmospheres of cool stars, through the simultaneous
observation of many clean CO, CN and OH lines.
A further advantage is the fact that in the 1-5$\,\mu$m domain,
lines from most molecules are often pure 
vibration-rotational lines.\footnote{Note, however, that this is not
the case for CN and C$_2$} The forest of molecular lines is also 
cleaner in the sense that several electronic systems less often
overlap severely compared to in the ultraviolet. Since the transitions within the vibration-rotational bands
occur within the electronic ground-state, the assumption of Local
Thermodynamic\linebreak Equilibrium (LTE) in the analysis of the molecules is
probably valid \citep{hinkle_lambert:75}. The assumption of
LTE simplifies the analysis considerably and should make it
more accurate.

Measurements of isotopic abundances of carbon, nitrogen, and
oxygen are of great interest, for example, for the study of
nucleosynthetic processes in stars and stellar evolution. The
infrared wavelength region is ideal for studying isotopic
abundances, chiefly since the isotopic shifts are larger for
molecules than for atoms, and molecular lines are ubiquitous in the
infrared. (The mass difference of two molecules of different isotopic mass leads to
unequal rotational constants, affecting the  wavelengths of
rotational and vibration-rotational lines of the
molecules.) For example, the isotopic shift between the
$^{12}$CO($v=1-3$) and $^{13}$CO($v=1-3$) band heads at $2.4\,\mu$m
is as much as $0.05\,\mu$m.

Several diagnostic lines of elements are stronger in the near-IR
beyond approximately $0.8\,\mu$m. For instance, the\linebreak  carbon
abundances in disk stars can be investigated from the lines at
$710$~nm \citep{tomkin} but a better diagnostic is the forbidden
line at 872.7 nm \citep{kol,bensby:kol}. For halo stars these
diagnostics are, however, too weak. Instead, the lines at $920$~nm
are much more useful. Also, when it comes to the origin of sulphur
in the halo, earlier lines at 869.4~nm \citep[for
instance][]{israel,takeda} have been used at large telescopes, but
the lines at 920~nm are approximately 10 times as strong
\citep{PEN:04,ryde:04_S,caffau,korn_ryde:05,ryde:05_S_poster} and should therefore be used. The
latter provide more accurate sulphur abundances and are also
detectable at 3 meter class telescopes provided these have
high-resolution spectrometers capable of detecting light up to
$1\,\mu$m. Recently, also sulphur lines beyond $1\,\mu$m have been
analysed, such as the [SI] line at 1082~nm \citep{ryde:forbidden:s},
and the lines at 1047~nm \citep{pen_crires}. These new lines are
providing new insight into the origin of sulphur. The near-IR is
preferred for a number of other elements too, e.g. the diagnostics
of HF, $^{13}$C, and Na  are superior in the IR.

The opacity in a cool, stellar atmosphere
has its minimum at $1.6\,\mu$m which implies that the continuum  is formed 
deepest in the atmosphere (where the physical state is relatively well known) at these wavelengths.
Furthermore, the intensity is less sensitive to temperature variations in the IR. Just as for black-body radiation
($B_\nu(T)$), $\delta B_\nu(T) / \delta T$ is small in the Rayleigh-Jeans  regime, which means that the
effects of uncertainties of, for instance, $T_{\mathrm{eff}}$ or surface inhomogeneities on
line strengths may be smaller in the IR \citep{ryde_munchen_review}.

\subsection{Drawbacks  of near-IR spectroscopy for abundance determinations}

The near-IR spectral region has not been used much previously for abundance analyses of stars. The optical wavelength region is much more studied. For instance, there are larger data bases of well-determined and well-studied optical spectral lines to use. Also, the use of optical spectra and photometry to determine the stellar parameters ($T_{\mathrm{eff}}$, $\log g$, metallicity, and microturbulence) are better developed in the optical regime. Howerver, as near-IR spectroscopy gets more commonly used and more work is done, these drawbacks will be alleviated. A few further drawbacks in doing an abundance analysis in the IR compared to the optical are discussed here.

There is  a large need for identifications, measured\linebreak   atomic wavelengths, and strengths (gf values) of atomic\linebreak  lines in the near-IR. The inventory of useful spectral lines has not been done, and  for the many lines which could potentially be very useful, line data (such as accurate wavelengths and transition probabilities which are  needed) are not known accurately enough or at all (see the review by \citet{sej:05}). Many lines from high-excitation levels in neutral and singly ionized atoms in the first half of the periodic table are present in the near-IR, which is actually also true for resonance lines of rare-earth elements.

Accurate wavelengths are possible to measure for all permitted lines, thanks to Fourier Transform Spectrometers (FTS) working in the near-IR.  Absolute gf values can be obtained by combining experimental lifetimes and branching fractions \citep{sej:05}. Note that for a relative abundance analysis, astrophysical gf values can also be used.  Very little experimental data for line strengths exists today. Theoretical data only exist for some lines. For example, theoretical  line strengths for hydrogenic transitions  are very accurate \citep{sej:05}. For neutral elements, the hydrogenic $n=4$ to $n=5$ transitions lie blue-wards of Br$\alpha$ at 4.05$\,\mu$m, and for singly ionized atoms these transitions lie close to $1\,\mu$m. The $4f-5g$ transitions are the most hydrogenic ones.  Such transitions are seen in the solar spectrum for  C, Na, Mg, Al, Si, Fe, Cr, and Ni \citep{geller}. For the alkali atoms ($3p-4s$ in Na{\sc i}  and $4s-4p$ in K{\sc i}) and alkaline earth elements (Mg{\sc i} and Ca{\sc i}), measurements of absolute gf values should be possible  \citep{sej:05}. This is also true for Iron-Group elements ($3d$ shell) which have hundreds of lines in the near-IR due to their complex atomic structures. For the Rare-Earth elements ($4f$ shell) it should also be possible to measure absolute gf values, but it will be difficult \citep{sej:05}. These are the elements (e.g. Ce{\sc iii}) with resonance lines in the near-IR. For the P-shell elements (e.g. the important C, N, O, Si, and S elements) it will be very difficult to measure lifetimes and branching ratios, and therefore one will need to rely on theoretical data or astrophysical gf values, with the uncertainties that come with these.


A further specific difficulty for an abundance analysis in the near-IR of IR bright stars is that several of the interesting ones have dust surrounding them. This dust radiates thermally in the
$2-10\,\mu$m region, making an abundance analysis very difficult if the amount of dust emission is not known.
Low-resolution spectra or a Spectral Energy Distribution is needed to estimate the dust contribution to the continuum level. Also circumstellar molecular layers seen in some cool giants \citep[see for example][]{tsuji:03} can  distort the classical picture. Line  profiles and  line centers should be carefully studied in these cases. 

Finally, the most efficient near-IR, high-resolution spectrometers today are single order echelle spectrometers. The wavelength coverage
is relatively small leading to a limited\linebreak  amount of spectral lines to be analyzed in a given amount of telescope time. 
A cross-dispersed spectrometer is needed to\linebreak  make near-IR abundance studies more efficient.

\subsection{The need for high spectral resolution}

For the purpose of this paper, I  define high spectral resolution as
$20,000\lesssim R = \lambda/\Delta\lambda \lesssim 60,000$, and
$R\gtrsim60,000$ very high resolution. Furthermore, we consider
$5,000\lesssim R\lesssim 20,000$ to be medium spectral resolution.
The lower limit is arbitrary but indicates a reasonable lower limit
for studying abundance indicators in stars. Thus the spectral
resolutions covered by the JWST are considered `low resolution'. From
medium resolution spectra, a metallicity indicator of a star can be
retrieved from measuring the strength of the near-IR Ca{\sc ii}
triplet at 0.86 \mic\ \citep[see for instance,][ and references
therein]{battaglia:2006,tolstoy:ca}. Furthermore, molecular bands can be
observed at medium resolution and especially isotopic ratios can be
retrieved (cf. \citet{isaac_delaverny}; \citet{shetrone}). However,
intrinsically, the line-broadening mechanisms (thermal, collision,
and micro- and macroturbulent broadening) in a stellar atmosphere
range from a few to tens of km\,s$^{-1}$ in velocity space. This
corresponds to a broadening of $\Delta\lambda\sim\lambda/20,000$ or
$\lambda/150,000$, which means that in order to resolve lines from
the stellar atmosphere we would need spectral resolutions of
$R=\lambda/\Delta\lambda\sim 20,000-150,000$.  Typical spectral resolutions for accurate stellar\linebreak 
abundance work is $R>40,000$, which is necessary for an accurate analysis of
stellar spectra
(see, for instance, Ryde et al. 2005).  Note, that for interstellar spectroscopy
a very high spectral resolution is crucial.

High, and very-high resolution spectroscopy in the near-IR have several advantages
\citep[see, for instance,][]{black} which should be taken into
account in the practical trade-offs. For example, there will be an
increased usefulness of the near-IR wavelength region, since only by
matching the resolution to the widths of the telluric
features, can these be corrected for properly.  Also, it will
enable the study of weak and narrow lines. Observations will be more sensitive to weak lines, which is advantageous at abundance analyses since their strengths are more sensitive to the abundances and less sensitive to other circumstances such as atmospheric motions and other line-broadening mechanisms. Generally, it is
obvious that more information can be retrieved concerning line
shifts and broadening, when working at very high spectral resolution.
Furthermore, the ubiquitous molecules that exist in the near-IR can
be studied in detail at high spectral resolution only,
resolving rotational and vibrational bands. As always, when
going towards\linebreak higher degree of detail we will certainly also
discover new features not realized before. For example,
going towards very high resolution ($R=80,000$), mid-infrared spectra of
warm, red super-giants for the first time
\citep{aboo,ryde:water3,ryde:water1,ryde:water2}, unexpected
spectral features were discovered in these well-studied stars. The
features, due to photospheric water vapor, were previously
overlooked since\linebreak earlier observations were made at lower spectral
resolution. The unexpected discovery of photo\-spheric water vapor\linebreak
pointed to a lack of understanding of the outer parts of the
atmospheres of red supergiants.

\subsection{The status of the modeling of stellar atmospheres}

In order to determine elemental abundances in stars, a\linebreak model of the stellar atmosphere is needed.
Subsequently a synthetic spectrum is computed. A discussion on the how realistic these steps are for cool stars is given in \citet{ryde_munchen_review}. It should be noted that the largest systematic uncertainties and problems in abundance analyses are mostly within the modeling of the stellar atmospheres and the calculation of synthetic spectra.

Several ingredients in the modeling procedure have uncertainties. First, there are uncertainties in the input parameters to the stellar atmosphere modeling  ($T_{\mathrm{eff}}$, $\log g$, metallicity, mass or radius and microturbulence). Second, in the modeling itself, several assumptions may be wrong or the physics may be wrongly described. The physical structures of the atmospheres depend on the accuracy of the opacities used. For cool giants, opacity from water vapor is uncertain. The situation is even more severe in model atmospheres of carbon stars, where the lack of accurate molecular opacities of  C$_2$H$_2$,
CH$_4$, C$_2$H, and C$_3$H hampers the modeling efforts. Other effects like sphericity of giants and supergiants \citep{plez:92}, non-LTE effects \citep{short:03}, inhomogeneities in the atmospheres \citep{asplund_3D} all have to be taken in to account to a certain level of realism. Furthermore,  the dynamic behavior of the atmosphere of supergiants \citep{freytag:2003,collet:07}, and mira stars \citep{bowen,hofner:97,hofner:03} has to be taken into account, both in the spatial structure of the atmosphere and its temporal evolution. For some stars the structures depart markedly from the static structure calculated by static models. For M dwarfs (and cooler)  and  mira stars, dust plays an important role in driving a stellar wind and shaping the model structure.  There can be an intricate interplay between radiation and dust in the outer parts of the atmospheres, which has to be modelled \citep[see, for example,][]{hofner:97}.  
Finally, magnetic fields on, for example, cool dwarfs could affect their structures, both vertically and horizontally, and should therefore be modelled. Also, magnetically-sensitive lines in the IR  are affected by magnetic fields through the Zeeman effect. 

Still only one-dimensional (1D), LTE models, treating convection crudely, exist as a standard for abundance analyses of red giants and dwarfs. Diagnostic tests of their validity are needed, and the development of 3-D, non-LTE models is needed to estimate  systematic errors in studies using standard models. An example of an unexpected discovery, which was made since the outer structure of the stellar atmosphere was not correctly described,  was the existence of atmospheric water-vapor in Arcturus, a star assumed to be too warm to have water in its atmosphere \citep{aboo,ryde:water3}. This discovery was based on water-vapor  lines in the mid-IR.  Furthermore, \citet{tsuji_1997,tsuji:03}  present empirical evidence of a molecular-forming region (MOLsphere) close to the photosphere of several M giants, which is neither expected theoretically. Clearly, caution has to be taken, and we can obviously not trust current  models in all cases.

For the calculation of synthetic spectra, line identifications, accurate wavelengths, transitions probabilities, statistical weights, excitation energies, together with line-\linebreak broadening parameters, partition functions, and dissociation energies (for molecules)  are needed.  The accuracy of the calculation depends on these input data and on the validity of the assumed approximations  of the physics.  For the IR, more data is still needed, both for atoms and a large number of molecules \citep{ryde_munchen_review}. If  a spectral line in study can not be assumed to be formed in  Local Thermodynamic Equilibrium, a full statistical-equilibrium calculation for the entire atom is needed. One  problem for such calculations is that the needed data for collisional transitions is uncertain or non-existent, which could leading to uncertain results.





\section{Examples of published studies}

The refereed literature on chemical
abundance analyses of cool stars based on medium and high-resolution\linebreak 
spectroscopy at $0.8-2.4\,\mu$m shows that this is clearly an emerging
field. We  can expect the field to generate much more scientific return
in the future, especially in the \elt\ era. Thanks to the
realization of high-resolution, infrared ($1-5\,\mu$m) spectrometers
such as the CRIRES\footnote{CRIRES  is the last of
the first generation of VLT instruments. It is more sensitive and provides a wider wavelength range than any other corresponding spectrometer available.} spectrometer \citep{crires1,crires2} on the VLT,
this field  will grow rapidly in the near future too.


\begin{figure*}
\includegraphics[width=170mm,height=100mm]{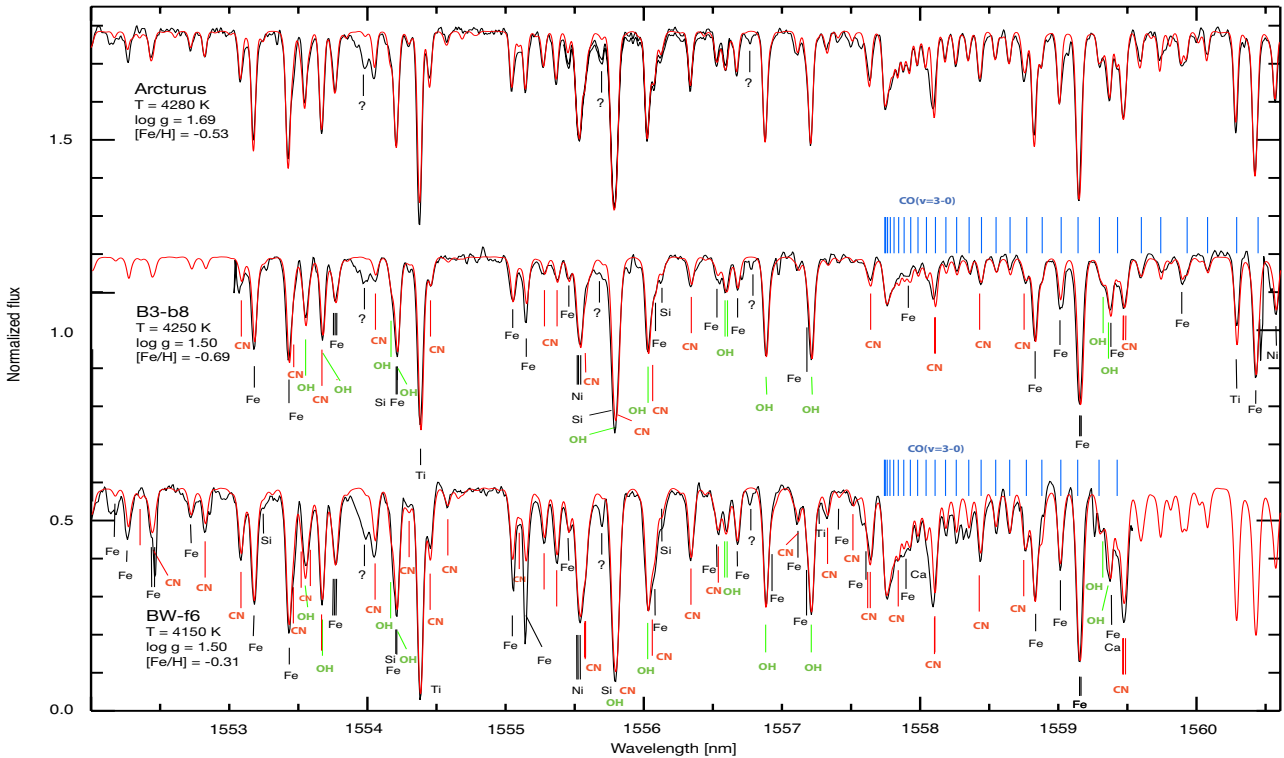}
\caption{Examples of near-IR spectra recorded with the CRIRES spectrometer of two giants (B3-b8 and BW-f6) in the bulge are shown with black lines. Models are shown in red. The spectra are of very high quality \citep[see ][]{ryde_bulb2}. The identified molecular and atomic lines are marked. On top is shown the $\alpha$ Boo spectrum \citep{arcturusatlas_II}.}
\label{label1}
\end{figure*}

An example of abundance analyses of cool stars at high spectral resolution is the work by \cite{pen_crires} who discuss the chemical evolution of sulphur, based in part on CRIRES data, but chiefly on the near-IR lines at $8694$ and $ 9213-38\,$ \AA\ observed with the UVES spectrograph.
Sulphur is of interest since it is not depleted on dust, which means
that sulphur is a good probe of chemical enrichment  and star-formation histories at
cosmological distances \citep{PEN:04}. 
In a number of projects on different telescopes, near-IR lines 
are now being used 
to increase the number of diagnostics available \citep{caffau,pen_crires,ryde:forbidden:s, korn_ryde:05,ryde:05_S_poster}. 
Another example is an ongoing CRIRES project \citep{ryde_crires, ryde_bg:08,ryde_bulb1,ryde:IAU09,ryde_rio,ryde_bulb2},  where the goal is to reveal the secrets of the Galactic bulge and 
to answer the questions how it was formed and how it
evolved,  benefitting from the dust penetration
properties of IR light \citep{cardelli}. Currently spectra of mainly bright giant stars in low-obscured regions are reachable (Fig. \ref{label1}), but in the \elt\ era the bulge will be exploitable in larger detail and in more regions.  Note, however, that recently microlensed bulge dwarfs have also been targetted \citep{bensby_micro}. \citet{melendez:2003} were the first to use high-resolution ($R=50,000$), near-IR spectroscopy of giants
in a globular cluster in the bulge (NGC6553). They performed a detailed abundance analysis of 5 giants using
CO (for the C abundance), CN (for N), and OH (for O) lines in the H band using the Phoenix spectrometer. 
\cite{cunha:2006} observed seven K and M giants in the bulge and determined C, N, O, Ti, Na, and Fe abundances from H and K spectra recorded with Phoenix. The C and N abundances show signatures of CN cycling and this implies that the oxygen abundance has not been altered during the stars' life time.
In general, the CNO abundances as well as the 
isotopic abundances of C and O in giants in the K band  are useful as probes of mixing
processes in red giants and are valuable in testing evolutionary
models. Predicted values are still not consistent with observed
values, which may imply that extra mixing is needed in the models.
Livia Origlia and co-workers
\citep[e.g.][]{origlia_GC4,origlia:08}
have written a number of papers on abundance studies of red giants,
both in the field and in open and globular clusters in the Galactic
bulge, based on NIRSPEC spectra\footnote{The NIRSPEC spectrometer \citep{nirspec} is one of the Keck II
telescope's most used spectrometers \citep{nirspec_mclean}. It is a
cross-dispersed echelle spectrometer, working at $0.96-5.5\,\mu$m at
resolutions of $R=2000-37,000$. It has a high-resolution mode
($R\sim20,000-30,000$) which has been used in studies on spectral
features and stellar abundances in, for example, brown dwarfs and
red giants and dwarfs.} which are at lower resolution. Furthermore, \cite{rich:2007}  presented the first detailed abundance analysis of 17 M giants 
of the \textit{inner} bulge, also based on NIRSPEC spectra.

The study by \cite{mcsaveney:2007} is a further example. They  determine abundances in luminous, intermediate-mass AGB stars in the Magellanic Clouds with the Phoenix spectrometer.  Their result is the first confirmation of a large production of primary nitrogen in these stars. Similarly,\linebreak \cite{wahlin:2005,wahlin:2006} used both high-resolution Phoenix spectra and medium-resolution ISAAC\footnote{ISAAC is the medium-resolution, near-IR spectrograph. It is able to  reach $R=10,000$, and it covers a larger wavelength range than  CRIRES.} spectra to determine the CNO abundances, and the $^{12}\mathrm C/^{13}\mathrm C$ ratio for carbon stars in the dwarf galaxies surrounding the Milky Way of different metallicities. Eventhough observations in the near-IR is already being performed of the close satellite galaxies of the Milky Way, with the E-ELT many more\linebreak galaxies will be targeted.
The optical high-resolution spectrograph UVES\footnote{UVES has a wavelength coverage in the red up to approximately $1\,\mu$m. The FLAMES spectrograph is a multi-object spectrograph capable of medium and high spectral resolution. In the GIRAFF configuration, 130 objects at a resolution of $R<25,000$ can be observed
and  when it is connected to UVES,  $R<47,000$ is achievable for a maximum of 8 objects.}  has  been used extensively for the determination of stellar abundances of galactic and extra-galactic stellar populations, the latter being a key science case of the \elt\ and which can be address by infrared spectra of longer wavelengths too. For example, \cite{kaufer:2004} determine the first  [Fe/H] and [$\alpha$/Fe] abundance ratios in the Dwarf Irregular Galaxy Sextans A, by observing three A type supergiants. They find a near solar  [$\alpha$/Fe], which is consistent with a slow chemical enrichment in these systems. Several investigations of dwarf spheroidal (dSph)\linebreak galaxies\footnote{In general, dSph galaxies are often seen as fossils from the early Universe and as building blocks of larger galaxies \citep{helmi:2006}. They are likely the most common galaxies in the Universe.} surrounding the Milky Way have been\linebreak performed, for example, by \cite{tolstoy:2003} and \linebreak \cite{shetrone:2003} using UVES, and\linebreak \cite{shetrone:2001} using HIRES at slightly lower resolution at KECK I.  Especially by the DART team \citep[Dwarf galaxies Abundances and Radial-velocities Team; ][]{tolstoy:2006}, using the multi-object capability of the\linebreak FLAMES spectrograph, have investigated the dSphs. Only the closest of these galaxies can, however, be observed in detail with 8 meter telescopes \citep{tolstoy:2006}. For example,  \cite{letarte:2007} make an abundance analysis of such spectra of 80 red giants in the 
Fornax dSph galaxy, showing that its star-formation was relatively slow. Furthermore, \cite{battaglia:2007} and \citet{helmi:2006} show the usefulness of, and use the Calcium triplet (CaT) in medium-resolution  spectra from red giants in dwarf\linebreak spheroidals to determine their metallically distributions.\linebreak \cite{tolstoy:2004} and \cite{battaglia:2006} show, for example,  evidence of several distinct stellar components in the Sculptor, and Fornax dSph galaxies, respectively. \cite{helmi:2006} show that, based on observations of  four dSphs, the progenitors to the dSphs are fundamentally different\linebreak  from the early components of the Milky Way. \cite{letarte:2006} observed nine stars in globular clusters in the Fornax dSph with UVES, showing how important it is to do spectroscopy at high spectral resolution. The conclusion for these works is that the abundances, spatial distributions, and the kinematics are not easy to interpret \citep{tolstoy:2006}.





As a final example, \citet{mclean} present a sequence of NIRSPEC spectra in the J band of 16 M, L, and T dwarfs (M2.5 to T6) in a
survey of brown dwarfs. They discuss the relative behavior of
spectral features through the sequence of stars and brown dwarfs. \citet{burgo} determine physical parameters of T dwarfs based on NIRSPEC spectra, and find, as expected, that T dwarfs later than T5 have dust free atmospheres due to dust sedimentation.

\section{Exploring stellar abundances based on near-IR observations at the \elt}

\subsection{Science drivers}

Science drivers are important science cases that are used to derive the requirements for the 
\elt\ telescope and its instruments.
A reference science driver for the \elt\ is the imaging (photometry)
and spectroscopy of resolved stellar populations \citep{swg}, which
aims at answering the fundamental question of the origin and
evolution of galaxies. This field represents one of the most
outstanding scientific challenges in modern astronomy and will
dominate galactic astrophysics for decades to come (see e.g.
Bland-Hawthorn \& Freeman, 2006 \nocite{bland} and Renzini,
2006\nocite{renzini}).
As yet there is no detailed understanding of galactic structure and
evolution that is physically consistent, even for the closest
galaxies (in the Local Group). For example, the number of faint
galaxies is below that theoretically predicted (the missing satellite
problem or cosmological substructure problem). Furthermore,
environmental effects have been\linebreak difficult to constrain, as also the
relative importance of internal dynamic processes in undisturbed
galaxies \citep{kormendy}. Indeed, it is not clear how the Milky Way
fits into the currently emerging cosmological model in detail. Stars
are being formed today but at low rates.
But the questions are when did the stars in 
central bulges of spiral galaxies form? Did they form early on or
later through merger-induced star formation? Obviously, we have to
learn more about the assembly sequence of the major components of
galaxies and the role of internal versus external processes in
building galaxies. The way forward is to collect more empirical
facts from observations in order to constrain theoretical models.
One way to elucidate such questions 
is to study the detailed chemical abundances of stars in large
numbers of stars in various populations and fields of galaxies (see
for example Silk \& Wyse, 1993\nocite{silk}).
This can be done by medium and high resolution stellar spectroscopy.


Hence, the study of resolved stellar populations of\linebreak galaxies in
order to understand the origin and evolution of galaxies, and the
formation of stars, is one of three highlighted science cases for
the \elt\ \citep{swg}. The spatial resolution of an ELT will allow
an exploration of individual stars even in galaxies beyond the 42
galaxies in the Local Group. Stars in several galaxies in the Virgo
Cluster (18 Mpc) should be observable by photometry providing
first-order ages and metallicities \citep{swg,tolstoy,momsi} and
stars in galaxies of the Local Group (within approximately 1 Mpc)
should be attainable spectroscopically, cf. Section \ref{spec:elt}.
All projects on ELT-sized telescope concepts today have the
formation and evolution of galaxies as one of a few key science
drivers. Among the six science cases for the GMT, we find `Stellar
Populations and Chemical Evolution' and 'the Evolution of Galaxies',
and the TMT pushes it as one of its drivers and highlights it as one
of the most ambitious scientific goals of the next decade.
Furthermore, for the JWST the `Assemblies of Galaxies' is one of
four science drivers. This field is also one of four targeted for
the long-term strategic planning of European Astrophysics as
expressed in the
Science Vision of Astronet.

\subsection{The spectrographic performance of an \elt.}
\label{spec:elt}

A general consideration in the planning of the \elt\ and its
instrumentation, is the balance sought in performance and
complementarity to ALMA and especially to the planned JWST. The goal
is to create synergy effects similar to those with the VLT/HST
combination, with follow-up observations. The area in the parameter
space that the \elt\ occupies by itself compared to JWST is the
better achievable spatial resolution, the larger collecting power,
and the possibility to carry out medium and high resolution spectroscopy. The
difference in telescope sizes of JWST (6 meters) and the E-ELT (42
m), means that not only will the spatial resolution of the \elt\ at
the diffraction limit be a factor of 7 better, but the \elt\ will
also collect 50 times more photons per second (corresponding to more
than 4 magnitudes). The JWST wins on the ability to detect weak
sources due to its higher sensitivity and the lower background in
space. The spectrometers planned (which are not restricted to
telluric windows) for the JWST will all have spectral resolution of less
than $R=2700$. The \elt\ has no principle restrictions on the
resolution for spectroscopic studies.

The detailed exploration of stellar abundances requires\linebreak spectroscopy
at high resolution ($R=\lambda/\Delta \lambda\gtrsim 40,000$).
More photons are needed in spectroscopy than for imaging and
photometry, which means that it will not be possible to observe as
faint stars as by photometry.  To estimate a relevant limiting
magnitude for spectroscopy (see Table \ref{lim}), we have estimated
the faintest magnitude of a star with which it is possible to
achieve a signal-to-noise ratio (SNR) per resolution element of approximately 50 for an
observation with an exposure time of 3 hours. This limit is of
course arbitrary and, depending on the science that is to be done,
the SNR may need to be higher. However, with a 42 m telescope,
observing at a high spectral resolution of $R=50,000$, and using laser
tomography AO, 
$I\sim19.0$, and $J=H=K\sim19.5$ is attainable. 
This calculation is based on the E-ELT Exposure Time Calculator in Spectroscopy Mode Version 2.14. Thus, based on the magnitudes in Table
\ref{mag}, Turn-Off (TO) stars in the Galactic bulge (but hardly
in the Large Magellanic Cloud), red giants in open and globular
clusters, in the Galactic bulge, and in dwarf spheroidal galaxies
surrounding the Milky Way (marginally in M31\footnote{RGB stars have
large ranges in bolometric magnitude depending on their position on
the giant branches. Observing red giants higher up on the branch
will make it possible to reach red giants in M31}) should be within
reach for an abundance analysis in the K band, even in dense fields. Also, red
giants at the Tip of the Red Giant Branch (TRGB) and supergiants
will be marginally observable out to the Sculptor Group, a group of galaxies
that contains several large spiral galaxies. Furthermore, the high
spatial resolution of a 42 m telescope will be crucial for crowded
star fields, such as the nuclei of Local Group galaxies such as the
Andromeda galaxy. However, in order to explore large elliptical
galaxies it is necessary to observe Cen A or galaxies in the Leo
group \citep[Tolstoy,][]{swg}. They are still too far away to make it possible to 
study stellar abundances in detail in elliptical
galaxies at high spectral resolution.




A measurement of the rough metallicity of a system can be obtained
at lower resolution ($R=5000-8000$) by observing the near-IR Ca{\sc
ii} triplet at 0.86 \mic\ \citep[see for example][]{tolstoy:ca}.
Also, isotopic ratios derived from molecular bands can be determined
from lower resolution spectra \citep{shetrone}. Thus, lowering the
resolution to $R \gtrsim 5,000$ implies reaching, for example, $K=23$ at a SNR per resolution element of
15,\footnote{For instance \citet{battaglia:2006} require a SNR per
resolution element of larger than 10 for an accurate measurement,
and \citet{bosler} measure calcium abundances from the Ca{\sc ii}
triplet in spectra of SNR per pixel greater than 10.} after 3 hours, see Table
\ref{lim}.
This means that red giant stars in all galaxies in the Local Group
can be analysed, and marginally in the galaxies in the Sculptor group.
Furthermore, TO stars in some of the dwarf galaxies surrounding the Milky
Way can be analysed. TRGB stars will be possible to analyse out to
M81/82\footnote{M82 is the nearest starburst galaxy with ongoing
star formation in a Super Star Cluster (SSC), cf. \cite{momsi}.} and
even out to Cen A.

\begin{table}
     \caption[]{Limiting magnitude estimates based on the ETC of the \elt ,  version 2.14. For the given resolution the approximate magnitudes
     are given for an observation with an exposure time of 3 hours achieving a SNR$\sim50$ per resolution element for the
     $R=50,000$ mode and SNR$\sim15$ per resolution element for the $R=5,000$ mode.  }
        \label{lim}
        \begin{tabular}{lcccc}
           \hline
       Resolution     & $I$ & $J$ & $H$ & $K$ \\
       \hline

$R=50,000$    & $19$ & $19.5$  & $19.5$   & $19.5$ \\
$R=5,000$     & $23.5$   & $24$    & $24$ & $23$ \\

       \hline

        \end{tabular}
\end{table}

\begin{table*}
     \caption[]{Distances, distance moduli, spatial resolutions, and magnitudes of red giants and turn-off stars in
     galaxies and galaxy components relevant for high-resolution, near-IR spectroscopy with the E-ELT.}
        \label{mag}
        \begin{tabular}{lcccccccc}
           \hline
           \noalign{\smallskip}
       Object      & Approx. distance & (m-M)$_0$ & $\Theta(1\,\,\mathrm{pc})$ & $I_\mathrm{TRGB}^{\mathrm{a}}$ & $K_\mathrm{TRGB}^{\mathrm{a}}$ & $K_\mathrm{RGB}^{\mathrm{a}}$ & $K_\mathrm{TO}^{\mathrm{a}}$\\
& $\mathrm{[Mpc]}$ & & & & \\
            \hline
           \noalign{\smallskip}           \noalign{\smallskip}

Galactic\,\,bulge   & 0.008 & $14.4^{\mathrm{b}}$ & 26"   & 10.5 & 8 & 12 & 17.5\\
LMC               & 0.05 & $18.5^{\mathrm{c}}$ & 4"   & 14.5 & 12 & 16 & 21.5\\
SMC               & 0.06 & $18.9^{\mathrm{c}}$ & 3.5"   & 15 & 12.5 & 16.5 & 22\\
Sculptor\,\,Dwarf\,\,Galaxy          & 0.09 & $19.7^{\mathrm{d}}$ & 2.3"   & 15.5 & 13 & 17 & 22.5\\
Fornax\,\,Dwarf\,\,Galaxy            & 0.14 & $20.7^{\mathrm{e}}$ & 1.5"   & 16.5 & 14 & 18 & 23.5\\
Leo\,\,I\,\,Dwarf\,\,Galaxy             & 0.25 & $22.0^{\mathrm{f}}$ & 0.8"   & 18 & 15.5 & 19.5 & 25\\

M31               & 0.7  & 24.3 & 0.3" & 20.5 
& 18
& 22 & \\

Local Group &  $\sim1.0$  & 25.0 & 0.2" & 21 & 18.5 & 22.5\\
Sculptor Group & 2.5  & 26.5 & 0.1" & 22.5 & 20 & 24\\
M81/82               & $\sim3.5$  & 27.8 & 0.06"& 24 
& 21.5 
& 25.5\\
Cen A          & $\sim3.5$  & 28.5 & 0.04"& 24.5 & 22 & \\
Virgo Cluster     & 18  & 30.9 & 0.014"&  & 24.5 \\

       \hline

        \end{tabular}
\begin{list}{}{}
\item[$^{\mathrm{a}}$] TRGB (tip of the RGB) and AGB: $M_V\sim -2.5$, $M_I\sim -4.0$
and $M_K\sim -6.5$; RGB stars: $M_V\sim -0.5$ and $M_K\sim -2.5$;
oldest turnoff stars, TO: $M_V\sim +4.5$ and $M_K\sim +3.0$
\citep{hook,arnold}.
\item[$^{\mathrm{b}}$] \citet{bulge}
\item[$^{\mathrm{c}}$] \citet{keller} 
\item[$^{\mathrm{d}}$] \citet{sculptor}
\item[$^{\mathrm{e}}$] \citet{fornax} 
\item[$^{\mathrm{f}}$]  \citet{bellazzini:2004}

\end{list}

\end{table*}

\subsection{Examples of scientific questions to be addressed}

There are, in principle, three different types of possible
applications at the \elt\ for determining stellar element\linebreak abundances
(Ryde et al. 2005). {\it First}, one could study\linebreak  fainter\footnote{in principle,
nearly 4 magnitudes fainter, only considering the size of the
telescope} objects than before, for example, reaching for dimmer,
nearby dwarfs or stars in external galaxies. For distant giants in
the Local Group galaxies, the analyses will be relatively
straight-forward as long as the stars seem similar to stars known
already from the Galaxy or the Magellanic Clouds. When chemical
abundances, or other spectral characteristics, seem different or
exotic -- in many respects the most interesting case -- the accuracy
in the atmospheric parameters derived will depend on the possibility
of the \elt\ to acquire high S/N and high-resolution spectra across
wide spectral regions. {\it Second}, one could study more objects in a given time, performing systematic studies of
populations or performing surveys of complete samples, e.g. within a
certain volume. The usefulness of surveys will be considerable as
they will presumably give further clues concerning the role of red
giants in stellar evolution and nucleosynthesis. {\it Third}, one
could strive towards higher accuracy in the observations or
the analyses, in order to achieve a better element-abundance analyses,
with details about observed atmospheric velocity-fields or magnetic
fields.

Examples of specific questions to address, by exploring\linebreak near-IR
spectra at high spectral resolution, are given here in random order. The \elt\ will
be able to address these questions for many more and fainter stars
in new environments and other galaxies. A large variety of new
populations will be investigated and intercompared.

\subsubsection{Chemical evolution of galaxies and the study of stellar populations}

The quantity of metals in stellar systems and in the Universe as
a whole grows with time. The relative distribution of the abundances
of different chemical elements and their different growths with
time, provide information on the star-formation rate and initial-mass function of the systems but also on 
internal processes in stars and
their explosions, and on processes in stellar evolution. This is
possible since different elements are synthesized by different
processes and in stars of different masses. Thus, chemical evolution of galaxies and the study of stellar populations can be investigated in detail for
a large range of galaxies of different types and morphology. Models
can best be tested with detailed abundances of an ensemble of stars.
In general, the precise element compositions (e.g. C, N, O, Mg, and Ca) and
kinematics can be determined, sampling different stages of the
chemical enrichment of a stellar system. The metallicity
distributions and the distribution of populations within stellar
systems can be determined.  Stars with unaltered surface compositions can be used as probes of
the parent cloud from which they were formed and as such trace the
star-formation history and enrichment of the interstellar medium.
This is encoded in the stellar elemental abundances. The life times
of the stars used in this context are long, comparable to the age of
the universe, and can therefore measure the star-formation rate and
metallicity at different times. 

In general, studying abundances of elements in stars in 
a population is important for the following reasons.
{\it First}, the $\alpha$ elements (Mg, Si, S, Ca, and Ti) are important tracers of the
star-formation rate (SFR) of the stellar system or
population being studied. The SFR can be determined from the
functional behaviour of the [$\alpha$/Fe] ratio as a function of
metallicity. A faster enrichment due to a high SFR will keep the
curve at a high value also for high metallicities. Precise elemental
abundances are required from many lines. This is only possible at
high spectral resolution. The behaviour of the different $\alpha$
elements can also be investigated for different stellar systems.
{\it Second}, the initial-mass function (IMF) of a stellar system or
population can be determined from the
over-abundance of $\alpha$ elements relative to the scaled solar
value. A shallower IMF will increase the number $\alpha$-element
producing stars thus raising the over abundance.
{\it Third}, the CNO abundances are important for many reasons. Oxygen is
particularly important since accurate O abundances and O/Fe ratios
in an ensemble of stars will also set strong constraints on the star
formation history of a stellar population. The OH lines at 1.55 $\mu$m are very useful (see,
for example, \citep{melendez2}). The determination of carbon
abundances is needed for O determinations, since CO molecules bind
much O in cool stars. Optical estimates of the C abundances are
highly uncertain (in most cases only upper limits), whereas the CO
and OH lines at 1.55 and 2.33 $\mu$m will together easily provide C
and O abundances. C is also interesting in itself. The precise site
for the formation of C is still being debated \citep{carigi}. In order to
determine the carbon abundance it is possible to measure lines from
CH, CO, C$_2$, and CI. The near-IR also has the virtue that only
that region offers all indicators necessary for an accurate
determination of the important C-N-O molecular and atomic equilibria
in the atmospheres of cool stars, through the simultaneous
observation of many clean CO, CN and OH lines. Finally, numerous CN
lines in the H band will make it possible to determine much more
accurate nitrogen abundances than from optical lines. The C+N
abundances reflect the original composition of the star better than
the two individual elements. The C/N ratio will also map the degree
of mixing in the stars as a function of metallicity and luminosity.

Examples of more specific questions that can be tackled are,
{\it first},  the dynamic evolution of the Milky Way
that will be better understood by exploring the fine-structure of
abundance patterns of different stellar populations
\citep{bg:unevolved}. Studies at high accuracy of Galactic halo and
thick disk stars will provide clues to early galaxy evolution.
Hypotheses of the enrichment of large galaxies compared to dwarf
galaxies and the enrichment of the intergalactic medium by outflows
caused by stellar winds from stars and supernova explosions in dwarf
galaxies can by tested.  The chemical enrichment in low-mass
galaxies seems to be an episodic process, whereas it is more smooth
in larger galaxies. But when were the first stars formed? How many
periods of large star-formation have there been? We know that there
was a period of enhanced star formation at intermediate redshifts
($2<z<4$). Was this the first and dominating episode of star
formation in the Universe? Studies of elemental abundances will be
able to shed light on this sort of questions. The chemical
evolution of galaxies will depend on the location and environment in
which the stars are embedded. It seems that chemical evolution is a
local process, implying that we need to investigate the chemical
evolution of many more systems, such as the dwarf spheroidal
galaxies and M31. {\it Second}, a comprehensive study of the components of the Milky Way,
based on a solid statistical foundation, with a large sample of
stars observed, will be of fundamental importance of the study of
the formation and evolution of galaxies in general. A study of the
relations of the thin and thick disks, and especially the least
studied component, the highly obscured Galactic bulge, would be of
major interest: are the disk and bulge separated components? Are
bulges in general associated with the inner disks or reflect the
last major merger? If disks grew later after a merger-formation of a
bulge then small disks should be younger than bulges surrounded by
large disks. In this picture there should be strong similarities
between bulges and ellipticals. Ellipticals can be seen as bulges
that have not had time  (or are in a dense environment) to form a
disk surrounding them. Therefore, a study of the Milky Way bulge, the
closest bulge that we can investigate in detail and which near-IR
spectrometers on the \elt\ will readily reach, is of great interest
in connection with and complementary to the exploration of stellar
populations in the
elliptical galaxies also attainable with the \elt.
{\it Third}, studying the properties of the halos, and
the thick and thin disks and bulges of nearby galaxies would give
empirical evidence of the formation of these structures.
Disentangling the heating mechanisms, whether internal or external,
which heat up the thick disks will be possible by studying stellar
abundances in disk galaxies \citep{arnold}.

\subsubsection{Nucleosynthesis of different elements}

The detailed nucleosynthesis of different elements can be
investigated by studying specific resolved lines of these elements. Metals are formed in the nuclei of stars and in the end phases of
stars' lives, in AGB stars and in supernova (SNe) explosions. They are later expelled into the Interstellar Medium (ISM), and used in the formation of a new generation of stars.  The yields from different types of stars are uncertain but are needed in the chemical evolution models. Accurate abundance studies of field stars and
of stars in dwIrr galaxies will provide knowledge of the role of
different production sites for different nuclei
\citep{bg:unevolved}.    Also, the role of SNe and hypernovea in low-metallicity regions can be investigated. 
An example is the investigations of the origin of \emph{fluorine}
which is still being debated and can be investigated in the near-IR for a
large sample of stars with the \elt. \citet{meynet:00} proposed that
the WR winds are significant sources of F. Thus, measurements of F,
as well as C, provide a test of the WR wind models. The only F
criteria are IR lines from the HF molecule. The vib-rot line at 
$2.34\,\mu$m is suitable for this purpose. Furthermore, stellar evolution and structure can be studied by looking at
the change of specific elements, such as C, N, and O but also the s-elements
(Y, Zr, Ba, La, Sr, Ce,...) in an evolutionary sequence of stars of different parameters and in different galaxies. AGB stars are luminous and red, thus favoring near-IR spectroscopy. With the \elt\ they can be studied in all Local Group galaxies.

\subsubsection{Red giants, miras, and carbon stars}

Investigations of red giants, miras, and carbon stars is challenging but important from an stellar evolution point-of-view. The understanding
of their evolution, their structures and atmospheres are of interest.  A majority of all stars will go through the Asymptotic Giant Branch phase, a stage that is not well understood, but very important, not the least as regards the mass return they are responsible for through their massive mass-losses. Furthermore, we do not understand the AGB stars' evolution with increasing bolometric luminosity. Investigations of AGB stars in different populations in different Local Group galaxies would be needed. Today, we can barely reach carbon stars for an abundance analysis in the Magellanic Clouds \citep{wahlin:2005}. The Milky Way seem to have a mixed population of carbon Stars.  These stars have very complicated spectra with a large amount of molecular lines and bands, which means that near-IR observations are necessary in order to analyze them. Dust-enshrouded carbon stars can also be studied much easier in the near-IR. 

Red giants are important as probes  for
studies of the chemical evolution of galaxies since they are bright. Therefore, the abundances and their variation with their evolution is important to understand. For example, their internal mixing processes can be studied in more detail. 
The \emph{carbon and oxygen isotopic ratios} and the nitrogen abundances (e.g. from CN and NH lines) are crucial indicators of the evolutionary phase of red giants and of the efficiency of their interior mixing processes. For instance,
the low ratio observed for some red giants, such as the bulge-like
star Arcturus, is still not understood. Investigations of red giants (and carbon stars) in more galaxies of different characteristics will add
to the investigations on the mixing processes in these stars. Also the mass-loss process from red giants
is important to study, since it is not understood from first principles and plays an important role in the
lives of these stars and the expulsion of newly synthesized elements into the ISM.

\subsubsection{The coolest dwarfs} 

Large surveys of the coolest dwarf stars (such as the M, L, T, and brown dwarfs)
can be done, which is important for  the analysis and modeling of these stars. Spectral features of 
CO in the near-IR is a powerful diagnostic tool to analyze M dwarfs.

\subsubsection{Extremely metal-poor stars}

The systematic study of extremely metal-poor stars [Fe/H]$<-3$, providing information
on the first steps of chemical enrichment and of early
star-formation of the first stars, will be possible with the \elt.
It may also give insights into the signatures of individual
Supernova Type II explosions and thereby constraints on their yields,
which are still very uncertain. The very first stars can be searched
for in the Galactic bulge where they are expected to be, representing
the earliest phases of the formation of the Milky-way or even the
protogalaxy. Also, due to the high spatial resolution of an \elt\ in the near-IR, the study can extend to
other Local Group galaxies. Since the lines of extremely metal-poor stars are very
weak, very high S/N ratios are needed, a feature the \elt\ also will be
able to deliver.

\subsubsection{Zeeman splitting}

Due to the sensitivity of certain infrared lines to magnetic fields through the Zeeman effect, studies of stellar magnetic field should be possible. High signal-to-noise ratios are needed.
The ratio of the wavelength separation of the Zeeman splitting ($\Delta\lambda_\mathrm{B}\sim g\lambda^2B$) and the non-magnetic Doppler width
($\Delta\lambda_\mathrm{D}\sim\lambda$), is given by
$\Delta\lambda_\mathrm{B}/\Delta\lambda_\mathrm{D}\sim g\lambda B$ \citep{ryde:04_letter,ryde:04_mg}.
Since the ratio grows linearly with wavelength, it is easier to detect Zeeman
split lines in the infrared compared to the optical. 


\section{Options and Requirements on ELT instrumentation for
stellar abundance work}





In order to discuss the requirements for near-infrared,\linebreak medium and
high-resolution spectrometer concepts for the \elt, a few key
projects are proposed here. {\it First}, the exploration of the elemental abundances in the atmospheres of red giants, subgiants, or
even {\it TO stars in the Galactic bulge}. A survey of abundances
of bulge stars will determine how the bulge was formed and give
crucial insights into galaxy formation and evolution in general. TO
stars will be reached. However, subgiants and red giants are
spectroscopically more interesting in the near-IR.
{\it Second}, the study of the {\it TRGBs in M81/82 and Cen A}. The Tip of the Red Giant Branch (TRGB) will
be reached in these galaxies at low spectral
resolution. The TRGB in galaxies in the Virgo Cluster will, however, be difficult. A survey of the diagnostically important Ca{\sc ii}
Triplet in these galaxies 
would give
metallicities 
of stars in a  range of galaxy
types. This is one of the main science drivers of ELTs in general.
{\it Third},  the investigation of the {\it First Stars}, that is, extremely metal-poor
stars in our and other galaxies. These can be observed by the \elt\ at very high SNR in order
to detect very weak lines.
{\it Fourth}, the exploration of the {\it Galactic Chemical Evolution in M31} and in other
galaxies of the Local Group in the same way that has been done for
the Milky Way \citep{reddy,bensby:kol}. Different components of
the galaxies, such as the thick and thin disks and the halo, can be
explored. Previous work on our galaxy has changed our view of it
profoundly. We also believe that other galaxies may look different.
In Table \ref{key} the specifications of important parameters are
shown for the different key projects.

\begin{table*}
     \caption[]{Requirements for a near-infrared, high spectral resolution spectrometer for the \elt\ from the point
     of view of explorations of stellar elemental abundances in a few chosen key projects. The SNR given is relative estimates
     normalized to a `medium SNR' of 100. For the angular resolution, it is interesting to compare this with
     the diffraction limit of a 42 m telescope which would be slightly less than $0.015"$.}
        \label{key}
    $$
        \begin{tabular}{lcccccccc}
           \hline
            & Subgiants in the & & TRGB stars & & First & & Galactic Chemical \\
            &  Galactic bulge && in M81/82 and Cen A && stars && Evolution in M31 \\
       \hline

Angular resolution & $\lesssim  2"$  && $\lesssim 0.04"$   && $\lesssim 3"$  && $\lesssim0.3"$\\
Spectral range    &  $0.8-2.5\,\mu\mathrm m$   && $0.8-2.5\,\mu\mathrm m$    && $0.8-2.5\,\mu\mathrm m$ && $0.8-2.5\,\mu\mathrm m$ \\
Spectral resolution & $R>60,000$ && $R>5,000$ && $R>60,000$ && $R>60,000$ \\
SNR & high && low && very high && high \\
Spectral coverage & full && parts && full && full \\
\,\,\,\,\,\,\,\,in one exposure & & & \\
Spatial density$^a$ & $\sim 500\,\,\mathrm{stars/(')^2}$  &&  $\lesssim50\,\,\mathrm{stars/('')^2}$ && N/A && $\sim 1 \,\,\mathrm{star/('')^2}$ \\
       \hline

        \end{tabular}
    $$
\footnotesize{$^\mathrm a$  The numbers for the spatial density of
stars given in the Table are based on the assumption that the stars
are evenly spaced in projected distance and the mean projected
distance between stars is estimated to 0.1 pc for dense stellar
fields like the Galactic bulge and 1 pc for red giants in galaxy
fields. This would imply approximately 500 stars/arcminute$^2$ for
the Galactic bulge. However, a check with a few 2MASS fields of the
bulge shows an order of magnitude lower spatial density of
identified stars. 2MASS has a magnitude limit of $K\sim14$ which
means that one can assume that it has detected most bulge red giants
(the reddening is 1/10 \citep{cardelli} of what it is in the visual,
implying that $A^\mathrm{bulge}_\mathrm{max}(K)< 3$). This factor of
ten is therefore used to estimate the density of red giants in M31,
where the \elt\ will only detect red giants. The same factor,
although it ought to be higher, is also used for an estimate of an
upper limit of the surface density of red giants in Cen A.}
\end{table*}
\subsection{Requirements for a medium- to high-resolution spectrometer
for the \elt}

Thus the requirements for a medium- to high-resolution\linebreak  spectrometer
for the \elt\ based on the scientific questions outlined above are provided below.

- A \textbf{high spatial resolution} is needed for crowded fields in the Galactic bulge and in
Local Group galaxies. This is achieved with a 42 m telescope with
adaptive optics. As noted earlier, in the K band the diffraction
limit will be less than $0.015"$ for the \elt, implying that stars
lying 0.05 pc apart in projected distance in M31 will be resolved,
cf. Table \ref{mag}. Likewise, stars lying 1 pc apart on the sky in
galaxies in the Virgo cluster will be resolved. In a typical galaxy
field a projected distance is 1 pc but gets smaller towards the
centre of the galaxy. In dense stellar fields (such as the galactic
bulge and globular cluster) distances of approximately 0.1 pc can be
used as a guide line.


- The \textbf{spectral range} of interest for high spectral resolution spectroscopy which is not covered
by the CODEX concept for the optical region \citep{codex}, is
$0.8-5.0\,\mu$m. At least the J, H, and K bands will provide a
necessary wavelength region, for instance enabling the determination
of $\alpha$ elements and the CNO abundances. See Table
\ref{elements} to see examples of which lines are available in the
different bands.

- Exploring stellar abundances in detail requires a high  \textbf{spectral resolution}. Stellar spectral lines are typically a few km\,s$^{-1}$ broad, which means that in order to resolve them an $R\sim100,000$ is needed. This will minimise blends. However, depending on the stars investigated a  lower resolution might still be appropriate, but a resolution of at least $R>40,000$ is  desirable.

- The two high-resolution instruments existing today at 8-10 meter telescopes are
the Phoenix \citep{phoenix:2003} and CRIRES \citep{crires1,crires2}
spectrometers. These are single order instruments, which hampers their
efficiency. A \textbf{full coverage} of a band in one or a few
exposures would increase the efficiency and scientific return with a
large amount. Thus, a cross-dispersed spectrometer would be required
and would increase the instantaneous wavelength coverage up to 20
times compared to that of CRIRES. A full wavelength coverage would
be a new feature for high-resolution spectrometers working in the
near-IR.

- In order to reach beyond the galaxies of the Local Group, i.e. to probe galaxies
of a wider range of morphological types, by high-resolution
spectroscopy, a \textbf{multi-object spectrometer} would make
observing more efficient and\linebreak make it possible at all. If a large
number (10-50) of red giants could be observed at once, then long
exposures could be justified, allowing us to start probing stars in
galaxies beyond the Local Group. Another way to proceed is to co-add
observations of a large number of stars to the same type, which
would require a multi-fiber system.

- An image slicer would increase the amount of light that is recorded by the spectrograph from the AO image on
the narrow (high-resolution) slit. A fiber-fed spectrograph would also
solve this problem.

\begin{table}
     \caption[]{Examples of elements for which spectral lines exist in the near-IR, suitable for an abundance analysis, see for example \citep{melendez:JH}.  }
        \label{elements}
        \begin{tabular}{lc}
           \hline
       Band     & Spectral diagnostics\\
       \hline
$J$ band &
CN, Na,
Al, Mn, Si, Ti, Fe, Mg, Sr, etc.\\
 ($1.0 - 1.3\,\mu\mathrm m$) & \\
$H$ band  &
CO ($\Delta v =
3$), OH, Mg, Al, Si, etc.\\
($1.5 - 1.8\,\mu\mathrm m$) & \\
$K$ band& CO ($\Delta v = 2$), Na, Ca, Si, Al, Mg, Ti, etc.\\
 ($2.0 - 2.4\,\mu\mathrm m$) & \\
$L$ band & OH, SiO($\Delta v = 2$), Mg, Ca, etc.\\
 ($3.3 - 4.2\,\mu\mathrm m$) & \\
$M$ band  & CO($\Delta v = 1$), SiO($\Delta v
=2$), Al, etc.\\
($4.5 - 5.5\,\mu\mathrm m$) & \\
       \hline

        \end{tabular}
\end{table}
%

\subsection{Current  \elt\ instrument studies}


In the Phase B study of the \elt\ there are 8 detailed instrument and two post-focal AO studies  currently well under way, and performed by institutes or consortia of institues in ESO member states \citep{sandro:2008}. Relevant studies for stellar abundance work include the SIMPLE and OPTIMOS concepts.  
{\it SIMPLE} is a near-IR, high-resolution spectrograph concept, being studied by a consortium lead by the Principle Investigator Livia Origlia (INAF, Osservatorio Bologna). It will cover $0.8-2.5\,\mu$m at very high resolution, $R=100,000$, and may even go up to $R=150,000$. It will be a cross-dispersed, single object spectrometer. It is
highly optimised for single point-source spectroscopy with full
wavelength coverage.  This would be the best concept for stellar abundance studies, even though it will not provide a possibility of observing several stars simultaneously. 
The near-IR, high-resolution concept was called  HISPEC-IR in the FP6
Small Study and corresponds to instrument No. 10 in the `ELT WG\#2:
Instruments'-report \citep{wg2}. The high spectral resolution instrument concept is not considered to be
among the first-light instruments on the \elt\ since the 
diffraction-limited performance required may not be implemented from the
start and the science cases are somewhat specialized. {\it OPTIMOS} is a wide field visual MOS extending up to $1.4\,\mu$m and will have a high-resolution mode, providing $R=50,000$. 


Of the other instrument studies, the {\it HARMONI} \citep{harmoni} and {\it EAGLE} \citep{eagle} spectrometers provide a medium spectral resolution in the near-IR ($0.8-2.5\,\mu$m) \citep{sandro:2008,cunningham}. In the FP6 Small Study, the near-IR, medium resolution spectrometer concept was called MOMSI
\citep{momsi,momsi2} and approximately MOMFIS
from the OWL study. 
In the `ELT WG\#2: Instruments'-report \citep{wg2} these instrument
concepts correspond to instrument No. 6 (and maybe 7): `Multi-Object
NIR Spectrograph (high spatial\linebreak resolution)'. 
{\it HARMONI} (short for 
High Angular-\linebreak Resolution, Monolithic Optical and Near-infrared field\linebreak spectrograph)
is planned to be a single-field, wide-band\linebreak spectrometer with a limited field-of-view of  $10\times5"$ with a spectral resolution of $R=5000$ and up to $R=20,000$. It will include a multi-conjugate adaptive optics module to reach the diffraction limit in spatial resolution.  The  {\it EAGLE} spectrometer concept is a wide-field,  multi-object, multi-field
spectrometer-imager using many deployable IFUs and uses MOAO. It has  the
characteristics of two spectroscopic modes, one at a resolution of $R=4000$ and one at $R=10,000$.   
The field of view is larger, more than 5 minutes of arc. 
Furthermore, the
wavelength range is planned for the \emph{JHK} bands, with a goal of
also including the \emph{I} and \emph{Z} bands. 
This means that a wavelength
range of $800-2500$ nm is anticipated. 
In the low-resolution mode, a
full wavelength coverage in one of the bands is set as a goal. In
the medium-resolution mode, a coverage of $\Delta
\lambda/\lambda\sim1/20$ is planned in one exposure. 
Such a spectrometer is required for many science cases, prominently that
of the study of resolved stellar populations \citep{cunningham}.





\section{Conclusions}

In spite of the emerging character of near-IR spectroscopy and, for instance, the lack of spectroscopic data, the near-IR is a wavelength region where much scientific progress
can be made, also concerning the determination of stellar abundances. It
is preferred over the optical region in conjunction with crowded fields
since the diffraction limit of a large telescope will be achieved in the
near-IR first, and in conjunction with dust-obscured regions in the Universe, due to the
lower opacity of gas and dust in the near-IR. Red giants, the probes
that will enable us to study stellar abundances in detail in all
Local Group galaxies, are brightest in the near-IR and their spectra
show several advantages, thus making the analyses more accurate.
Furthermore, the important C, N, and O abundances are more
accurately determined in the near-IR, as are isotopic ratios from
molecular lines. However, the advantage of the near-IR depends on
the sensitivity of the IR detector arrays compared with the optical
CCDs. Also the possibility of a larger wavelength coverage gives
the optical an advantage, if the near-IR spectrometers are not
cross-dispersed. Depending on the
project, the near-IR would be complementary to the optical region.

For a detailed abundance analysis a high spectral resolution is
needed. Blends and uncertainties can only be minimized by observing
at high resolution. Especially red giants, late M-stars, and carbon stars can have messy spectra
that can only be disentangled by analyzing the intrinsic spectrum of
the stars, i.e. a stellar spectrum that is not degraded by an
instrumental profile that is much broader than the intrinsic widths
of the stellar features. However, some questions can be
addressed at medium resolution too, such as observing the metallicity
indicator (Ca {\sc ii} Triplet) and molecular bands.

The SIMPLE spectrometer concept, being studied for the E-ELT, will 
provide a high spectral resolution in the near-IR, which would allow a detailed abundance
analysis of stars in stellar populations at large distances.  The EAGLE \citep{eagle} and HARMONI \citep{harmoni} spectrometer concepts 
will satisfy the needs at medium
spectral resolution and the CODEX  \citep{codex} plans the needs for a high
spectral resolution spectrometer working in the optical
[$\lambda=0.4-0.8\,\mu$m range \citep{stc:66}].  

No doubt, with high-resolution spectrometers in the\linebreak near-IR at
extremely large telescopes, in combination with more realistic
models of cool stars, we may eventually look forward to uncovering
the detailed secrets of stellar populations and stellar evolution in galaxies in the Local
Group but also to reliable abundance determinations in a variety of
fundamentally interesting objects in the future. As a major side
effect, or perhaps as the most rewarding part of the effort, we will
learn a lot more about the physics of cool stars, the interesting
interplay between stellar nuclear reactions, pulsations and
traveling shocks, magnetic fields, convection, mass loss, dust
formation and non-equilibrium radiation fields.


\acknowledgements
This work was initiated while the author was
a visitor to the E-ELT Instrumentation Department at the ESO Garching headquarters in 2007 and 2008. 
The author wishes to thank Sandro D'Odorico for giving him the opportunity to write this paper and for reading and giving valuable comments improving the manuscript. 
The author also wishes to thank Bengt Gustafsson and Suzie Ramsay for carefully reading the manuscript.


%


\begin{thebibliography}{}

\bibitem[\protect\astroncite{{Arnold} et~al.}{2001}]{arnold}
{Arnold}, L., {Bacon}, R., {Davies}, R., et~al.: 2001,
\newblock in {{www.astro-opticon.org/fp5\ /leiden\_gal.ps}}

\bibitem[\protect\astroncite{{Asplund} et~al.}{2000}]{asplund_3D}
{Asplund}, M., {Nordlund}, {\AA}., {Trampedach}, R., {Allende Prieto}, C., and
  {Stein}, R.~F.: 2000,
\newblock {A\&A} {359}, 729

\bibitem[\protect\astroncite{{Battaglia} et~al.}{2008}]{battaglia:2007}
{Battaglia}, G., {Irwin}, M., {Tolstoy}, E., et~al.: 2008,
\newblock {MNRAS} {383}, 183

\bibitem[\protect\astroncite{{Battaglia} et~al.}{2006}]{battaglia:2006}
{Battaglia}, G., {Tolstoy}, E., {Helmi}, A., et~al.: 2006,
\newblock {A\&A} {459}, 423

\bibitem[\protect\astroncite{{Bellazzini} et~al.}{2004}]{bellazzini:2004}
{Bellazzini}, M., {Gennari}, N., {Ferraro}, F.~R., \& {Sollima}, A.: 2004,
\newblock {MNRAS} {354}, 708

\bibitem[\protect\astroncite{{Bensby} \& {Feltzing}}{2006}]{bensby:kol}
{Bensby}, T. \& {Feltzing}, S.: 2006,
\newblock {MNRAS} {367}, 1181

\bibitem[\protect\astroncite{{Bensby} et~al.}{2009}]{bensby_micro}
{Bensby}, T., {Feltzing}, S., {Johnson}, J. A., et al.: 2009,
\newblock {ArXiv} {0911.5076} 

\bibitem[\protect\astroncite{{Black}}{2005}]{black}
{Black}, J.~H.: 2005,
\newblock in H.~U. {K{\"a}ufl}, R. {Siebenmorgen}, \& A.~F.~M. {Moorwood}
  (eds.), {High Resolution Infrared Spectroscopy in Astronomy}, p.~3

\bibitem[\protect\astroncite{{Bland-Hawthorn} \& {Freeman}}{2006}]{bland}
{Bland-Hawthorn}, J. \& {Freeman}, K.~C.: 2006,
\newblock {Memorie della Societa Astronomica Italiana} {77}, 1095

\bibitem[\protect\astroncite{{Bosler} et~al.}{2007}]{bosler}
{Bosler}, T.~L., {Smecker-Hane}, T.~A., \& {Stetson}, P.~B.: 2007,
\newblock {MNRAS} {378}, 318

\bibitem[\protect\astroncite{{Bowen}}{1988}]{bowen}
{Bowen}, G.~H.: 1988,
\newblock {ApJ} {329}, 299

\bibitem[\protect\astroncite{{Caffau} et~al.}{2005}]{caffau}
{Caffau}, E., {Bonifacio}, P., {Faraggiana}, R., et~al.: 2005,
\newblock {A\&A} {441}, 533

\bibitem[\protect\astroncite{{Cardelli} et~al.}{1989}]{cardelli}
{Cardelli}, J.~A., {Clayton}, G.~C., \& {Mathis}, J.~S.: 1989,
\newblock {ApJ} {345}, 245

\bibitem[\protect\astroncite{{Carigi} et~al.}{2005}]{carigi}
{Carigi}, L., {Peimbert}, M., {Esteban}, C., \& {Garc{\'{\i}}a-Rojas}, J.:
  2005,
\newblock {ApJ} {623}, 213

\bibitem[\protect\astroncite{{Collet} et~al.}{2007}]{collet:07}
{Collet}, R., {Asplund}, M., \& {Trampedach}, R.: 2007,
\newblock {A\&A} {469}, 687

\bibitem[\protect\astroncite{{Connes}}{1970}]{connes}
{Connes}, P.: 1970,
\newblock {ARA\&A} {8}, 209

\bibitem[\protect\astroncite{{Crampton} et~al.}{2009}]{crampton}
{Crampton}, D., {Simard}, L., \& {Silva}, D.: 2009,
\newblock in A. {Moorwood} (ed.), {Science with the VLT in the ELT Era}, p. 279

\bibitem[\protect\astroncite{{Cuby} et~al.}{2008}]{eagle}
{Cuby}, J.-G., {Morris}, S., {Bryson}, I., et~al.: 2008,
\newblock in I.~S. {McLean} \& M.~M. {Casali} (eds.), {Ground-based and
  Airborne Instrumentation for Astronomy II.}, Vol. 7014 of {Proc. SPIE}, p.
  70141K

\bibitem[\protect\astroncite{{Cunha} \& {Smith}}{2006}]{cunha:2006}
{Cunha}, K. \& {Smith}, V.~V.: 2006,
\newblock {ApJ} {651}, 491

\bibitem[\protect\astroncite{{Cunningham} et~al.}{2006}]{wg2}
{Cunningham}, C., {D'Odorico}, S., {Kerber}, F., et~al.: 2006,
\newblock in {ELT Working Group \#2: Instrumnets, issue 1.3, 25.04.06}

\bibitem[\protect\astroncite{{Cunningham} et~al.}{2008}]{cunningham}
{Cunningham}, C., {Evans}, C., {Monnet}, G., \& {Le Louarn}, M.: 2008,
\newblock in T.~E. {Andersen} (ed.), {Extremely Large Telescopes: Which
  Wavelengths? Retirement Symposium for Arne Ardeberg.}, Vol. 6986 of {Proc.
  SPIE}, p. 69860K

\bibitem[\protect\astroncite{{Del Burgo} et~al.}{2009}]{burgo}
{Del Burgo}, C., {Mart{\'{\i}}n}, E.~L., {Zapatero Osorio}, M.~R., and
  {Hauschildt}, P.~H.: 2009,
\newblock {A\&A} {501}, 1059

\bibitem[\protect\astroncite{{D'Odorico}}{2008}]{sandro:2008}
{D'Odorico}, S.: 2008,
\newblock in F. {Bresolin}, P.~A. {Crowther}, \& J. {Puls} (eds.), {Massive
  Stars as Cosmic Engines}, Vol. 250 of {IAU Symposium}, p. 495

\bibitem[\protect\astroncite{{D'Odorico} et~al.}{2006}]{xshooter}
{D'Odorico}, S., {Dekker}, H., {Mazzoleni}, R., et~al.: 2006,
\newblock in I.~S. {McLean} \& M. {Iye} (eds.), {Ground-based and Airborne
  Instrumentation for Astronomy.}, Vol. 6269 of {Proc. SPIE}, p. 626933

\bibitem[\protect\astroncite{{Elias} et~al.}{2006}]{gnirs}
{Elias}, J.~H., {Joyce}, R.~R., {Liang}, M., et~al.: 2006,
\newblock in I.~S. {McLean} \& M. {Iye} (eds.), {Ground-based and Airborne
  Instrumentation for Astronomy.}, Vol. 6269 of {Proc. SPIE}, p. 62694C

\bibitem[\protect\astroncite{{ELT SWG}}{2006}]{hook}
{ELT SWG}: 2006,
\newblock in I. {Hook} (ed.), {The Science Case for the European Extremely
  Large Telescope: The next step in mankind's quest for the Universe}

\bibitem[\protect\astroncite{{ESO/STC-430}}{2007}]{stc:66}
{ESO/STC-430}, 5 April 2007

\bibitem[\protect\astroncite{{Evans} et~al.}{2006a}]{momsi}
{Evans}, C., {Atad}, E., {Hastings}, P., et~al.: 2006a,
\newblock in {{EC FP6: E-ELT Design Study: Momsi, a multi-object, multi-field
  spectrometer \& imager for the E-ELT.}}, Vol. ELT-TRE-UKA-11200-0002

\bibitem[\protect\astroncite{{Evans} et~al.}{2006b}]{momsi2}
{Evans}, C., {Cunningham}, C., {Atad-Ettedgui}, E., et~al.: 2006b,
\newblock in I.~S. {McLean} \& M. {Iye} (eds.), {Ground-based and Airborne
  Instrumentation for Astronomy.}, Vol. 6269 of {Proc. SPIE}, p. 62692V

\bibitem[\protect\astroncite{{Freytag} \& {H{\"o}fner}}{2003}]{freytag:2003}
{Freytag}, B. \& {H{\"o}fner}, S.: 2003,
\newblock {Astronomische Nachrichten Supplement} {324}, 173

\bibitem[\protect\astroncite{{Geller}}{1992}]{geller}
{Geller}, M.: 1992,
\newblock in {{Key to Identification of Solar Features, NASA Ref. Publ. 1224,
  vol III (Washington, DC) }}

\bibitem[\protect\astroncite{{Gullieuszik} et~al.}{2007}]{fornax}
{Gullieuszik}, M., {Held}, E.~V., {Rizzi}, L., et~al.: 2007,
\newblock {A\&A} {467}, 1025

\bibitem[\protect\astroncite{{Gustafsson}}{1999}]{bg:unevolved}
{Gustafsson}, B.: 1999,
\newblock in J.~R. {Walsh} \& M.~R. {Rosa} (eds.), {Chemical Evolution from
  Zero to High Redshift}, p.~1

\bibitem[\protect\astroncite{{Gustafsson} et~al.}{1999}]{kol}
{Gustafsson}, B., {Karlsson}, T., {Olsson}, E., {Edvardsson}, B., \& {Ryde},
  N.: 1999,
\newblock {A\&A} {342}, 426

\bibitem[\protect\astroncite{{Helmi} et~al.}{2006}]{helmi:2006}
{Helmi}, A., {Irwin}, M.~J., {Tolstoy}, E., et~al.: 2006,
\newblock {ApJ} {651}, L121

\bibitem[\protect\astroncite{{Hickson}}{2006}]{tmt-news}
{Hickson}, P.: 2006,
\newblock in {{TMT Newsletter Issue 6, Dec. 2006}}

\bibitem[\protect\astroncite{{Hinkle} et~al.}{1995}]{arcturusatlas_II}
{Hinkle}, K., {Wallace}, L., \& {Livingston}, W.~C.: 1995,
\newblock {{Infrared atlas of the Arcturus spectrum, 0.9-5.3 microns}},
\newblock San Francisco, Calif.~: Astronomical Society of the Pacific, 1995.

\bibitem[\protect\astroncite{{Hinkle} et~al.}{2003}]{phoenix:2003}
{Hinkle}, K.~H., {Blum}, R.~D., {Joyce}, R.~R., et~al.: 2003,
\newblock in P. {Guhathakurta} (ed.), {Discoveries and Research Prospects from
  6- to 10-Meter-Class Telescopes II.}, Vol. 4834 of {Proc. SPIE}, p. 353

\bibitem[\protect\astroncite{{Hinkle} et~al.}{1998}]{phoenix}
{Hinkle}, K.~H., {Cuberly}, R.~W., {Gaughan}, N.~A., et~al.: 1998,
\newblock {SPIE} {3354}, 810

\bibitem[\protect\astroncite{{Hinkle} \& {Lambert}}{1975}]{hinkle_lambert:75}
{Hinkle}, K.~H. \& {Lambert}, D.~L.: 1975,
\newblock {MNRAS} {170}, 447

\bibitem[\protect\astroncite{{H\"ofner} \& {Dorfi}}{1997}]{hofner:97}
{H\"ofner}, S. \& {Dorfi}, E.~A.: 1997,
\newblock {A\&A} {319}, 648

\bibitem[\protect\astroncite{{H{\"o}fner} et~al.}{2003}]{hofner:03}
{H{\"o}fner}, S., {Gautschy-Loidl}, R., {Aringer}, B., \& {J{\o}rgensen},
  U.~G.: 2003,
\newblock {A\&A} {399}, 589

\bibitem[\protect\astroncite{{Israelian} \& {Rebolo}}{2001}]{israel}
{Israelian}, G. \& {Rebolo}, R.: 2001,
\newblock {ApJ} {557}, L43

\bibitem[\protect\astroncite{{Johansson}}{2005}]{sej:05}
{Johansson}, S.: 2005,
\newblock in H.~U. {K{\"a}ufl}, R. {Siebenmorgen}, \& A. {Moorwood} (eds.),
  {High Resolution Infrared Spectroscopy in Astronomy}, p.~62

\bibitem[\protect\astroncite{{Kaluzny} et~al.}{1995}]{sculptor}
{Kaluzny}, J., {Kubiak}, M., {Szymanski}, M., et~al.: 1995,
\newblock {A\&AS} {112}, 407

\bibitem[\protect\astroncite{{Kashikawa} et~al.}{2002}]{focas}
{Kashikawa}, N., {Aoki}, K., {Asai}, R., et~al.: 2002,
\newblock {PASJ} {54}, 819

\bibitem[\protect\astroncite{{Kaufer} et~al.}{2004}]{kaufer:2004}
{Kaufer}, A., {Venn}, K.~A., {Tolstoy}, E., {Pinte}, C., \& {Kudritzki},
  R.-P.: 2004,
\newblock {AJ} {127}, 2723

\bibitem[\protect\astroncite{{K{\"a}ufl} et~al.}{2006}]{crires2}
{K{\"a}ufl}, H.~U., {Amico}, P., {Ballester}, P., et~al.: 2006,
\newblock {The Messenger} {126}, 32

\bibitem[\protect\astroncite{{Keller} \& {Wood}}{2006}]{keller}
{Keller}, S.~C. \& {Wood}, P.~R.: 2006,
\newblock {ApJ} {642}, 834

\bibitem[\protect\astroncite{{Kobayashi} et~al.}{2000}]{IRCS}
{Kobayashi}, N., {Tokunaga}, A.~T., {Terada}, H., et~al.: 2000,
\newblock in M. {Iye} \& A.~F. {Moorwood} (eds.), {Optical and IR Telescope
  Instrumentation \& Detectors}, Vol. 4008 of {Proc. SPIE}, p. 1056

\bibitem[\protect\astroncite{{Kormendy} \& {Kennicutt}}{2004}]{kormendy}
{Kormendy}, J. \& {Kennicutt}, Jr., R.~C.: 2004,
\newblock {ARA\&A} {42}, 603

\bibitem[\protect\astroncite{{Korn} \& {Ryde}}{2005}]{korn_ryde:05}
{Korn}, A.~J. \& {Ryde}, N.: 2005,
\newblock {A\&A} {443}, 1029

\bibitem[\protect\astroncite{{Lambert} et~al.}{1984}]{lambert:84}
{Lambert}, D.~L., {Brown}, J.~A., {Hinkle}, K.~H., \& {Johnson}, H.~R.: 1984,
\newblock {ApJ} {284}, 223

\bibitem[\protect\astroncite{{Lambert} et~al.}{1986}]{lambert:86}
{Lambert}, D.~L., {Gustafsson}, B., {Eriksson}, K., \& {Hinkle}, K.~H.: 1986,
\newblock {ApJSS} {62}, 373

\bibitem[\protect\astroncite{{Letarte} et~al.}{2006}]{letarte:2006}
{Letarte}, B., {Hill}, V., {Jablonka}, P., et~al.: 2006,
\newblock {A\&A} {453}, 547

\bibitem[\protect\astroncite{{Letarte} et~al.}{2007}]{letarte:2007}
{Letarte}, B., {Hill}, V., \& {Tolstoy}, E.: 2007,
\newblock in {EAS Publications Series}, Vol.~24 of {Engineering and Science},
  p.~33

\bibitem[\protect\astroncite{{Maillard}}{1978}]{maillard}
{Maillard}, J.~P.: 1978,
\newblock in M. {Hack} (ed.), {High resolution spectrometry}, p. 108

\bibitem[\protect\astroncite{{McLean}}{2005}]{nirspec_mclean}
{McLean}, I.~S.: 2005,
\newblock in H.~U. {K{\"a}ufl}, R. {Siebenmorgen}, \& A.~F.~M. {Moorwood}
  (eds.), {High Resolution Infrared Spectroscopy in Astronomy}, p.~25

\bibitem[\protect\astroncite{{McLean} et~al.}{1998}]{nirspec}
{McLean}, I.~S., {Becklin}, E.~E., {Bendiksen}, O., et~al.: 1998,
\newblock in A.~M. {Fowler} (ed.), {Infrared Astronomical Instrumentation},
  Vol. 3354 of {Proc. SPIE}, p. 566

\bibitem[\protect\astroncite{{McLean} et~al.}{2007}]{mclean}
{McLean}, I.~S., {Prato}, L., {McGovern}, M.~R., et~al.: 2007,
\newblock {ApJ} {658}, 1217

\bibitem[\protect\astroncite{{McSaveney} et~al.}{2007}]{mcsaveney:2007}
{McSaveney}, J.~A., {Wood}, P.~R., {Scholz}, M., {Lattanzio}, J.~C., and
  {Hinkle}, K.~H.: 2007,
\newblock {MNRAS} {378}, 1089

\bibitem[\protect\astroncite{{Mel{\'e}ndez} \& {Barbuy}}{1999}]{melendez:JH}
{Mel{\'e}ndez}, J. \& {Barbuy}, B.: 1999,
\newblock {ApJS} {124}, 527

\bibitem[\protect\astroncite{{Mel{\'e}ndez} et~al.}{2003}]{melendez:2003}
{Mel{\'e}ndez}, J., {Barbuy}, B., {Bica}, E., et~al.: 2003,
\newblock {A\&A} {411}, 417

\bibitem[\protect\astroncite{{Mel{\'e}ndez} et~al.}{2001}]{melendez2}
{Mel{\'e}ndez}, J., {Barbuy}, B., \& {Spite}, F.: 2001,
\newblock {ApJ} {556}, 858

\bibitem[\protect\astroncite{{Meynet} \& {Arnould}}{2000}]{meynet:00}
{Meynet}, G. \& {Arnould}, M.: 2000,
\newblock {A\&A} {355}, 176

\bibitem[\protect\astroncite{{Moorwood}}{2005}]{crires1}
{Moorwood}, A.: 2005,
\newblock in H.~U. {K{\"a}ufl}, R. {Siebenmorgen}, \& A.~F.~M. {Moorwood}
  (eds.), {High Resolution Infrared Spectroscopy in Astronomy}, p.~15

\bibitem[\protect\astroncite{{Nishiyama} et~al.}{2006}]{bulge}
{Nishiyama}, S., {Nagata}, T., {Sato}, S., et~al.: 2006,
\newblock {ApJ} {647}, 1093

\bibitem[\protect\astroncite{{Nissen} et~al.}{2007}]{pen_crires}
{Nissen}, P.~E., {Akerman}, C., {Asplund}, M., et~al.: 2007,
\newblock {A\&A} {469}, 319

\bibitem[\protect\astroncite{{Nissen} et~al.}{2004}]{PEN:04}
{Nissen}, P.~E., {Chen}, Y.~Q., {Asplund}, M., \& {Pettini}, M.: 2004,
\newblock {A\&A} {415}, 993

\bibitem[\protect\astroncite{{Noguchi} et~al.}{1998}]{hds}
{Noguchi}, K., {Ando}, H., {Izumiura}, H., et~al.: 1998,
\newblock in S. {D'Odorico} (ed.), {Optical Astronomical Instrumentation}, Vol.
  3355 of {Proc. SPIE}, p. 354

\bibitem[\protect\astroncite{{Olsen} et~al.}{2006}]{olsen:2006}
{Olsen}, K.~A.~G., {Blum}, R.~D., {Stephens}, A.~W., et~al.: 2006,
\newblock {AJ} {132}, 271

\bibitem[\protect\astroncite{{Origlia} et~al.}{2002}]{origlia_GC4}
{Origlia}, L., {Rich}, R.~M., \& {Castro}, S.: 2002,
\newblock {AJ} {123}, 1559

\bibitem[\protect\astroncite{{Origlia} et~al.}{2008}]{origlia:08}
{Origlia}, L., {Valenti}, E., \& {Rich}, R.~M.: 2008,
\newblock {MNRAS} {388}, 1419

\bibitem[\protect\astroncite{{Pasquini} et~al.}{2006}]{codex}
{Pasquini}, L., {Cristiani}, S., {Dekker}, H., et~al.: 2006,
\newblock in P. {Whitelock}, M. {Dennefeld}, \& B. {Leibundgut} (eds.), {The
  Scientific Requirements for Extremely Large Telescopes}, Vol. 232 of {IAU
  Symposium}, p. 193

\bibitem[\protect\astroncite{{Plez} et~al.}{1992}]{plez:92}
{Plez}, B., {Brett}, J.~M., \& {Nordlund}, {\AA}.: 1992,
\newblock {A\&A} {256}, 551

\bibitem[\protect\astroncite{{Ramsey} et~al.}{2003}]{HET:MRS}
{Ramsey}, L.~W., {Engel}, L.~G., {Sessions}, N., et~al.: 2003,
\newblock in M. {Iye} \& A.~F.~M. {Moorwood} (eds.), {Instrument Design and
  Performance for Optical/Infrared Ground-based Telescopes.}, Vol. 4841 of
  {Proc. SPIE}, p. 1036

\bibitem[\protect\astroncite{{Recio-Blanco} \& {de
  Laverny}}{2007}]{isaac_delaverny}
{Recio-Blanco}, A. \& {de Laverny}, P.: 2007,
\newblock {A\&A} {461}, 13

\bibitem[\protect\astroncite{{Reddy} et~al.}{2006}]{reddy}
{Reddy}, B.~E., {Lambert}, D.~L., \& {Allende Prieto}, C.: 2006,
\newblock {MNRAS} {367}, 1329

\bibitem[\protect\astroncite{{Renzini}}{2006}]{renzini}
{Renzini}, A.: 2006,
\newblock {ARA\&A} {44}, 141

\bibitem[\protect\astroncite{{Rich} et~al.}{2007}]{rich:2007}
{Rich}, R.~M., {Origlia}, L., \& {Valenti}, E.: 2007,
\newblock {ApJ} {665}, L119

\bibitem[\protect\astroncite{{Ryde}}{2006}]{ryde:forbidden:s}
{Ryde}, N.: 2006,
\newblock {A\&A} {455}, L13

\bibitem[\protect\astroncite{{Ryde}}{2008}]{ryde_bg:08}
{Ryde}, N.: 2008,
\newblock {Physica Scripta Volume T} {133(1)}, 014033

\bibitem[\protect\astroncite{{Ryde}}{2009a}]{ryde_rio}
{Ryde}, N.: 2009a,
\newblock {ArXiv e-prints} 0909.0135

\bibitem[\protect\astroncite{{Ryde}}{2009b}]{ryde:IAU09}
{Ryde}, N.: 2009b,
\newblock in J. {Andersen}, J. {Bland-Hawthorn}, \& B. {Nordstr{\"o}m} (eds.),
  {The Galaxy Disk in Cosmological Context}, Vol. 254 of {IAU Symposium}, p.
  159

\bibitem[\protect\astroncite{{Ryde} et~al.}{2009a}]{ryde_bulb1}
{Ryde}, N., {Edvardsson}, B., {Gustafsson}, B., et al.: 2009,
\newblock {A\&A} {496}, 701

\bibitem[\protect\astroncite{{Ryde} et~al.}{2007}]{ryde_crires}
{Ryde}, N., {Edvardsson}, B., {Gustafsson}, B., \& {K{\"a}ufl}, H.-U.: 2007,
\newblock in A. {Vazdekis} \& R.~F. {Peletier} (eds.), {Stellar Populations as
  Building Blocks of Galaxies}, Vol. 241 of {IAU Symposium}, p. 260

\bibitem[\protect\astroncite{{Ryde} et~al.}{2009b}]{ryde_bulb2}
{Ryde}, N., {Gustafsson}, B., {Edvardsson}, B., et al.: 2010,
\newblock {A\&A} {590}, A20 

\bibitem[\protect\astroncite{{Ryde} et~al.}{2005}]{ryde_munchen_review}
{Ryde}, N., {Gustafsson}, B., {Eriksson}, K., \& {Wahlin}, R.: 2005,
\newblock in H.~U. {K{\"a}ufl}, R. {Siebenmorgen}, \& A.~F.~M. {Moorwood}
  (eds.), {High Resolution Infrared Spectroscopy in Astronomy}, p. 365

\bibitem[\protect\astroncite{{Ryde} et~al.}{2006a}]{ryde:water2}
{Ryde}, N., {Harper}, G.~M., {Richter}, M.~J., {Greathouse}, T.~K., \& {Lacy},
  J.~H.: 2006a,
\newblock {ApJ} {637}, 1040

\bibitem[\protect\astroncite{{Ryde} et~al.}{2004}]{ryde:04_mg}
{Ryde}, N., {Korn}, A.~J., {Richter}, M.~J., \& {Ryde}, F.: 2004,
\newblock {ApJ} {617}, 551

\bibitem[\protect\astroncite{{Ryde} \& {Lambert}}{2004}]{ryde:04_S}
{Ryde}, N. \& {Lambert}, D.~L.: 2004,
\newblock {A\&A} {415}, 559

\bibitem[\protect\astroncite{{Ryde} \& {Lambert}}{2005}]{ryde:05_S_poster}
{Ryde}, N. \& {Lambert}, D.~L.: 2005,
\newblock in T.~G. {Barnes} \& F.~N. {Bash} (eds.), {ASP Conf. Ser. 336:
  Cosmic Abundances as Records of Stellar Evolution and Nucleosynthesis}, p.
  355

\bibitem[\protect\astroncite{{Ryde} et~al.}{2002a}]{aboo}
{Ryde}, N., {Lambert}, D.~L., {Richter}, M.~J., \& {Lacy}, J.~H.: 2002a,
\newblock {ApJ} {580}, 447


\bibitem[\protect\astroncite{{Ryde} et~al.}{2003}]{ryde:water3}
{Ryde}, N., {Lambert}, D.~L., {Richter}, M.~J., {Lacy}, J.~H., and
  {Greathouse}, T.~K.: 2003,
\newblock in S. {Turcotte}, S.~C. {Keller}, \& R.~M. {Cavallo} (eds.), {3D
  Stellar Evolution}, Vol. 293 of {Astronomical Society of the Pacific
  Conference Series}, p. 214

\bibitem[\protect\astroncite{{Ryde} \& {Richter}}{2004}]{ryde:04_letter}
{Ryde}, N. \& {Richter}, M.~J.: 2004,
\newblock {ApJ} {611}, L41

\bibitem[\protect\astroncite{{Ryde} et~al.}{2006b}]{ryde:water1}
{Ryde}, N., {Richter}, M.~J., {Harper}, G.~M., {Eriksson}, K., \& {Lambert},
  D.~L.: 2006b,
\newblock {ApJ} {645}, 652

\bibitem[\protect\astroncite{{Shetrone} et~al.}{2003}]{shetrone:2003}
{Shetrone}, M., {Venn}, K.~A., {Tolstoy}, E., et~al.: 2003,
\newblock {AJ} {125}, 684

\bibitem[\protect\astroncite{{Shetrone}}{2003}]{shetrone}
{Shetrone}, M.~D.: 2003,
\newblock {ApJ} {585}, L45

\bibitem[\protect\astroncite{{Shetrone} et~al.}{2001}]{shetrone:2001}
{Shetrone}, M.~D., {C{\^o}t{\'e}}, P., \& {Sargent}, W.~L.~W., 2001:
\newblock {ApJ} {548}, 592

\bibitem[\protect\astroncite{{Short} \& {Hauschildt}}{2003}]{short:03}
{Short}, C.~I. \& {Hauschildt}, P.~H.: 2003,
\newblock {ApJ} {596}, 501

\bibitem[\protect\astroncite{{Silk} \& {Wyse}}{1993}]{silk}
{Silk}, J. \& {Wyse}, R.~F.~G.: 1993,
\newblock {Phys. Rep.} {231}, 293

\bibitem[\protect\astroncite{{Spinrad} et~al.}{1971}]{spinrad}
{Spinrad}, H., {Kaplan}, L.~D., {Connes}, P., {Connes}, J., {Kunde}, V.~G., and
  {Maillard}, J.-P.: 1971,
\newblock {Contributions from the Kitt Peak National Observatory} {554}, 59

\bibitem[\protect\astroncite{{SWG report}}{2006}]{swg}
{SWG report}, 30 April 2006,
\newblock in {{Science Cases and Requirements for the ESO ELT. Report of the
  ELT Science Working Group}}

\bibitem[\protect\astroncite{{Takada-Hidai} et~al.}{2002}]{takeda}
{Takada-Hidai}, M., {Takeda}, Y., {Sato}, S., et~al.: 2002,
\newblock {ApJ} {573}, 614

\bibitem[\protect\astroncite{{Tecza} et~al.}{2009}]{harmoni}
{Tecza}, M., {Thatte}, N., {Clarke}, F., \& {Freeman}, D.: 2009,
\newblock in A. {Moorwood} (ed.), {Science with the VLT in the ELT Era}, p. 267

\bibitem[\protect\astroncite{{Tolstoy}}{2006}]{tolstoy}
{Tolstoy}, E.: 2006,
\newblock in P. {Whitelock}, M. {Dennefeld}, \& B. {Leibundgut} (eds.), {The
  Scientific Requirements for Extremely Large Telescopes}, Vol. 232 of {IAU
  Symposium}, p. 293

\bibitem[\protect\astroncite{{Tolstoy} et~al.}{2006}]{tolstoy:2006}
{Tolstoy}, E., {Hill}, V., {Irwin}, M., et~al.: 2006,
\newblock {The Messenger} {123}, 33

\bibitem[\protect\astroncite{{Tolstoy} et~al.}{2001}]{tolstoy:ca}
{Tolstoy}, E., {Irwin}, M.~J., {Cole}, A.~A., et~al.: 2001,
\newblock {MNRAS} {327}, 918

\bibitem[\protect\astroncite{{Tolstoy} et~al.}{2004}]{tolstoy:2004}
{Tolstoy}, E., {Irwin}, M.~J., {Helmi}, A., et~al.: 2004,
\newblock {ApJ} {617}, L119

\bibitem[\protect\astroncite{{Tolstoy} et~al.}{2003}]{tolstoy:2003}
{Tolstoy}, E., {Venn}, K.~A., {Shetrone}, M., et~al.: 2003,
\newblock {AJ} {125}, 707

\bibitem[\protect\astroncite{{Tomkin} et~al.}{1995}]{tomkin}
{Tomkin}, J., {Woolf}, V.~M., {Lambert}, D.~L., \& {Lemke}, M.: 1995,
\newblock {AJ} {109}, 2204

\bibitem[\protect\astroncite{{Tsuji}}{2003}]{tsuji:03}
{Tsuji}, T.: 2003,
\newblock in {ESA SP-511: Exploiting the ISO Data Archive. Infrared Astronomy
  in the Internet Age}, p.~93

\bibitem[\protect\astroncite{{Tsuji} et~al.}{1997}]{tsuji_1997}
{Tsuji}, T., {Ohnaka}, K., {Aoki}, W., \& {Yamamura}, I.: 1997,
\newblock {A\&A} {320}, L1

\bibitem[\protect\astroncite{{Tull}}{1998}]{HET:HRS}
{Tull}, R.~G.: 1998,
\newblock in S. {D'Odorico} (ed.), {Optical Astronomical Instrumentation}, Vol.
  3355 of {Proc. SPIE}, p. 387

\bibitem[\protect\astroncite{{Vogt} et~al.}{1994}]{hires}
{Vogt}, S.~S., {Allen}, S.~L., {Bigelow}, B.~C., et~al.: 1994,
\newblock in D.~L. {Crawford} \& E.~R. {Craine} (eds.), {Instrumentation in
  Astronomy VIII.}, Vol. 2198 of {Proc. SPIE}, p. 362

\bibitem[\protect\astroncite{{Wahlin} et~al.}{2005}]{wahlin:2005}
{Wahlin}, R., {Eriksson}, K., {Gustafsson}, B., et~al.: 2005,
\newblock in H.~U. {K{\"a}ufl}, R. {Siebenmorgen}, \& A.~F.~M. {Moorwood}
  (eds.), {High Resolution Infrared Spectroscopy in Astronomy}, p. 439

\bibitem[\protect\astroncite{{Wahlin} et~al.}{2006}]{wahlin:2006}
{Wahlin}, R., {Eriksson}, K., {Gustafsson}, B., et~al.: 2006,
\newblock {Memorie della Societa Astronomica Italiana} {77}, 955

\end{thebibliography}

\appendix

\section{Other suitable instruments for abundance studies in the near-IR}


There are a number of other suitable spectrometers in operation which are worth mentioning. For example, the Hobby-Eberly telescope's high and medium resolution
spectrographs (HET\linebreak MRS/HRS) which provid $R=$ 30,000-120,000 up to
1100~nm \citep{HET:HRS} and $R=$ 5,000-10,000 up to $1800$~nm
\citep{HET:MRS}, respectively. The HIRES spectrometer \citep{hires} at KECK I has a red configuration working up to $1\,\mu$m with resolutions
of $R=25,000-85,000$. GNIRS, the Gemini South telescope's near-IR Spectrograph \citep{gnirs} works in the near-IR
($1-5.5\,\mu$m) at a resolution up to $R=18,000$. IRCS,  the Subaru telescope's medium resolution echelle spectrometer\linebreak \citep{IRCS} covers
the spectral region of $0.96-5.6\,\mu$m and, finally, FOCAS \citep{focas} and HDS \citep{hds} also at Subaru covers a spectral range to $1\,\mu$m
at up to $R=7500$ and $100,000$, respectively.

There also exist  instrumentation plans for
existing 8-10 meter class telescopes.\footnote {In passing it
can be noted that for the JWST (scheduled to be launched in 2013)
the most relevant spectrometer planned is the \emph{NIRSPEC}
spectrometer
which works at low resolution ($R_\mathrm{max}=3000$). }
The VLT 2$^\mathrm{nd}$ generation instruments are planned for the
period 2008-2012. \emph{Xshooter} has recently been\linebreak commissioned and was highly oversubscribed in the first  call for observations (Spring 2009). It will be able to observe from the UV to the
K-band in one observation at medium resolution \citep{xshooter}. In
the visual-red arm (550-1000 nm) it will achieve a spectral
resolution of 12,000 and in the near-IR  arm (1000-2500 nm) it will
have a $R\sim 7500$ for a 0.6" slit. Limiting AB magnitudes, for a 1
hour exposure and S/N=10 per resolution element, are $I=21.2$,
$J=20.5$, $H=20.8$, and $K'=19.3$. The \emph{KMOS} instrument will work from
1000-2500 nm and have $R= 3400-3800$, i.e. low resolution.
At GEMINI the plans for the \emph{HRNIRS}, a near-infrared
high-resolution spectrometer, are unclear. It was chosen in 2003 for
further review, but was not endorsed as it is at the Aspen meeting in 2005.
Finally, at the Large Binocular Telescope, LBT, the \emph{Pepsi}
spectrometer will be able to achieve resolutions of
$R=40,000-300,000$ up to 1050~nm and the near-IR spectrometer,
LUCIFER, will achieve resolutions up
to $R=40,000$.

Apart from the \elt\ there are two other ELT concepts under study today. One of
them is the Thirty Meter Telescope, TMT, which is being planned  as a
collaboration between USA, Canada and Japan (and is the result of the merging of the earlier CELT, GSMT,
and VLOT ELT efforts). It will have a 30m, highly segmented primary
mirror consisting of 738 individual 1.2-meter mirrors, and is
scheduled for first-light in 2018. The other ELT concept is the
Giant Magellan Telescope, GMT, which is an USA, Australian and Korean
initiative to build an ELT. It will consist of 7 individual 8m
primary mirror segments having a resolving power corresponding to a
24.5 meter telescope, and a light collecting power corresponding to
a 21.2 meter telescope. It too has a scheduled completion date of 2018.

TMT instrument plans \citep{crampton} are divided into `early-light' and
`first-decade' instruments. Three early-light instruments, which are
planned to be operational from the start, have been chosen
\citep{tmt-news}. These are, {\it first}, 
the IR imaging spectrometer, \emph{IRIS}, performing diffraction-limited
spectroscopy with slits or multiple employable IFU in the  $0.8-2.5$
\mic\ region at a resolution of $R=4,000$. This resolution is
marginally what we have defined as medium resolution. One of three
science cases for this spectrometer is the study of stellar populations
in galaxies out to the Virgo Cluster.
{\it Second}, the IR multi-slit spectrometer, \emph{IRMS} which is a clone of
MOSFIRE for KECK ($0.9-2.5\,\mu$m, $R_\mathrm{max}=5,000$ with full
coverage.).
{\it Third}, the wide-field optical spectrometer, \emph{WFOS}, which is a multi-object spectrometer
mainly for the region up to $1.1\,\mu$m, but it once had the goal of
reaching 1.6 \mic. The spectral resolution planned for is $R\lesssim
5,000$. The goal is still to achieve $R\lesssim 7,500$

`First-decade' instruments, relevant for stellar spectroscopy, which
are planned to be operational within the first decade of the TMT's
life-time are, (in order of decreasing spectral resolution), {\it first}
the very relevant spectrometer  \emph{NIRES}, which is a near-infrared Echelle
Spectrograph. It will be a diffraction-limited spectrometer for the
$1-2.4$\,\mic\ region providing high-resolution spectra of
$R=20,000-100,000$.
{\it Second},  \emph{HROS}, the High-Resolution Optical Spectrometer, will work up to
at least $1\,\mu$m (with the goal of reaching up to $1.3\,\mu$m) at a
resolution of $50,000<R<90,000$. {\it Finally}, \emph{IRMOS}, the Infrared Multi-Object Spectrograph, will work in the
near-infrared ($0.8-2.5$\,\mic) at a resolution of $R=2,000-10,000$
with multiple Integral Field Units (IFU) to access a 5' field.

The Giant Magellan Telescope (GMT) consortium has identified a few
instrument concepts which are candidates for the set of first
generation instruments to be developed. Relevant for stellar spectroscopy  in the near-IR 
is the GMTNIRS: the GMT near-IR High-Resolution Spectrometer. It is
planned to work at $1-5$\,\mic, and provide spectra at a resolution
of $R\sim 50,000-120,000$.

\end{document}